\documentclass[11pt, preprint]{aastex}   		
\usepackage{geometry}                		
\usepackage{graphicx}				
								
\usepackage{amssymb}

\usepackage{booktabs}
\usepackage{lscape}

\shorttitle{UV ISM Diagnostics: Metallicity and Dust}
\shortauthors{Zetterlund et al.}
\slugcomment{\today}     

\begin{document}
\title{Ultraviolet ISM Diagnostics for Star-Forming Galaxies I. Tracers of Metallicity and Extinction} 
\author{Erika Zetterlund$^1$, Emily M. Levesque$^{1,3}$, Claus Leitherer$^2$, Charles W. Danforth$^1$}
\footnotetext[1]{CASA, Department of Astrophysical and Planetary Sciences, University of Colorado 389-UCB, Boulder, CO 80309, USA; contact: \texttt{Erika.Zetterlund@colorado.edu}}
\footnotetext[2]{Space Telescope Science Institute, 3700 San Martin Drive, Baltimore, MD 21218, USA}
\footnotetext[3]{Hubble Fellow}

\begin{abstract}
We have observed a sample of 14 nearby ($z \sim 0.03$) star-forming blue compact galaxies in the rest-frame far-UV ($\sim1150-2200$ \AA) using the Cosmic Origins Spectrograph on the Hubble Space Telescope. We have also generated a grid of stellar population synthesis models using the Starburst99 evolutionary synthesis code, allowing us to compare observations and theoretical predictions for the SiIV\_1400 and CIV\_1550 UV indices; both are comprised of a blend of stellar wind and interstellar lines and have been proposed as metallicity diagnostics in the UV. Our models and observations both demonstrate that there is a positive linear correlation with metallicity for both indices, and we find generally good agreement between our observations and the predictions of the Starburst99 models (with the models slightly under-estimating the value of the indices due to contributions from interstellar lines not simulated by a stellar population synthesis code). By combining the rest-frame UV observations with pre-existing rest-frame optical spectrophotometry of our blue compact galaxy sample, we also directly compare the predictions of metallicity and extinction diagnostics across both wavelength regimes. This comparison reveals a correlation between the UV absorption and optical strong-line diagnostics, offering the first means of directly comparing ISM properties determined across different rest-frame regimes. Finally, using our Starburst99 model grid we determine theoretical values for the short-wavelength UV continuum slope, $\beta_{18}$, that can be used for determining extinction in rest-frame UV spectra of star-forming galaxies. We consider the implications of these results and discuss future work aimed at parameterizing these and other environmental diagnostics in the UV (a suite of diagnostics that could offer particular utility in the study of star-forming galaxies at high redshift) as well as the development of robust comparisons between ISM diagnostics across a broad wavelength baseline.  
\end{abstract}

\section{Introduction}

A galaxy's redshift plays a dominant role in determining the rest-frame wavelength bands in which it is most easily observed. The rest-frame UV spectra of local galaxies can be observed with UV-sensitive instruments. The rest-frame UV spectra of galaxies at $z \sim 2-3$, on the other hand, must be observed in the optical. Such optical observations offer many practical advantages over observing the more distant galaxies in the IR, in order to access their rest-frame optical spectra, due in large part to an abundance of strong night sky emission lines in the IR. With $z \sim 2-3$ galaxies sampling the regime when star formation density peaks \citep{Hopkins06.1}, it is therefore advantageous to utilize UV data from local starburst galaxies to develop spectral interstellar medium (ISM) diagnostics that can ultimately be applied to optical observations of higher-redshift galaxies.

Historically, the development of diagnostics for determining oxygen-abundance metallicity (log(O/H) + 12) in star-forming galaxies has largely been confined to the optical. The ``direct" metallicity diagnostic uses the auroral [OIII] $\lambda$4363 line to determine the electron temperature, $T_e$, of the gas in galaxy, which can then be converted to a metallicity (e.g. \citealt{Aller84.1}). However, this auroral line is weak even at low metallicities and this method is therefore rarely usable in high-metallicity or faint galaxies. As a result, diagnostics utilizing observed strong-line ratios have been calibrated against $T_e$ metallicities (e.g. \citealt{Pettini04.1}). Strong-line diagnostics have also been developed based on theoretical calibrations using population synthesis models along with photoionization models (e.g. \citealt{Kobulnicky04.1}), or through a combination of observational and theoretical methods (e.g. \citealt{Denicolo02.1}). Unfortunately, the numerous metallicity diagnostics available in the optical \citep{Kobulnicky04.1,Pettini04.1} have been found to show poor agreement, although conversions do exist which facilitate comparisons between different methods \citep{Kewley08.1}.

Fewer diagnostics are available in the UV. Interstellar absorption lines in the UV are produced by the resonance transitions of abundant ions. These lines are stronger in starbursts, due to the larger interstellar velocity dispersions. The stronger interstellar lines tend to be optically thick and therefore their equivalent widths depend more on velocity dispersion than column density \citep{GonzalezDelgado98.1}. Because the more metal-rich galaxies tend to host more powerful starbursts, those galaxies tend to have more energy in the ISM. This increases the velocity dispersion and therefore the equivalent widths of the interstellar lines \citep{Leitherer11.1}. This metallicity dependency has been shown by \citet{Heckman98.1} and \citet{Leitherer11.1}. Stellar photospheric absorption lines can also be used as metallicity diagnostics when the spectra have the high signal-to-noise and moderate resolution required to measure equivalent widths of $\sim 1$ \AA\ \citep{Leitherer11.1}. To aid in this resolution problem, \citet{Rix04.1} employed the indices defined by \citet{Leitherer01.1} to determine that a region between 1935 \AA\ and 2020 \AA, containing many Fe \textsc{iii} photospheric lines, scales with metallicity.

Stellar wind lines --- resonance lines produced in the radiatively driven winds of O and B stars \citep{Leitherer11.1} --- offer the most robust means of directly tracing metallicity in the UV. The lines originate mainly from gas at $T \gtrsim 10^4$K which has been ionized by radiation from massive stars along with collisional processes in the outflow \citep{Shapley03.1}. The winds are driven by photon momentum transfer through metal-line absorption, so the stellar wind lines are metallicity-dependent \citep{Leitherer11.1}.  \citet{Leitherer11.1} developed a set of indices, similar to the Lick indices \citep{Burstein84.1,Faber85.1}, for the purpose of measuring interstellar and stellar wind lines in the UV. These indices measure the equivalent width of blended absorption lines in a systematic manner. This allows for the consistent measuring of line strengths even in spectra where the resolution is not high enough to measure individual line widths.

Unfortunately, robust conversions between UV and optical metallicity diagnostics are non-existent. The optical and UV light of star-forming galaxies originate from different processes --- UV stellar spectra consist mainly of the light from O and B stars of young stellar populations, while optical stellar spectra trace older stellar populations \citep{Rix04.1} --- making a clear relationship between diagnostics in the two regimes even more critical.

A similar difficulty exists for spectral diagnostics of dust extinction in star-forming regions. The effective extinction is defined as the difference in magnitudes of the attenuated light and the dust-free source \citep{Calzetti01.1}, and sensitive diagnostics are required to determining the intrinsic attenuation due to dust. In the optical, the Balmer decrement (H$\alpha$/H$\beta$) is most commonly used to quantify extinction effects. The intrinsic Balmer decrement is expected to be 2.86 assuming case B recombination ($T_e = 10^4$ K and $n_e \sim 10^2 - 10^4$ cm$^{-2}$, following \citealt{Osterbrock89.1}), with a higher observed decrement indicating increased dust content. When UV and optical spectra are both available, line-to-continuum ratios H$\alpha$/UV and H$\beta$/UV are also popular \citep{Calzetti01.1}.

The slope of the UV continuum is another widely-used extinction diagnostic. The UV spectral slope, $\beta$ is determined by fitting the continuum to the function $f(\lambda) \propto \lambda^\beta$, where $f(\lambda)$ is the flux density, in a specified wavelength range. The wavelength range 1250--2600\AA\ (which defines $\beta_{26}$) is most commonly used \citep{Calzetti01.1}. The UV spectral slope traces the reddening effects of dust on the continuum. Dust is proportional to metallicity and gas content, which evolve in opposite directions with time. Whereas metallicity increases with time as more stars have formed and produced metals, more gas is locked away in stars as time goes by. Thus the dust column density peaks at some point within the galaxy's evolution, with an overall peak predicted at a redshift of $z \sim 2$ \citep{Calzetti99.1}.

However, there are degeneracies between extinction effects and other galaxy properties, including the age of the stellar population, which dominates the spectral energy distribution (SED); the effects of metallicity; and the initial mass function (IMF). These become especially troublesome when only low-resolution spectroscopy is available \citep{Papovich01.1}. Furthermore, reddening effects are sensitive to the geometry of the dust distribution when dealing with extended sources such as star-forming regions. Typically the light is not as attenuated as would be expected from a simple dust-screen model \citep{Witt92.1}. In addition, spectral coverage in the UV often spans a narrower wavelength range than that defined by the $\beta_{26}$ diagnostic, limiting its efficacy as an extinction diagnostic at higher redshifts.

In this paper we present a preliminary analysis of UV ISM diagnostics for metallicity and extinction. Using UV and optical spectra of nearby ($z \sim 0.03$) star-forming galaxies (Section 2) along with a grid of stellar population synthesis models (Section 3), we examine both observed and theoretical predictions for the utility of two potential UV diagnostics for metallicity and their correlations with optical strong-line diagnostics (Section 4). We also present our measurements of the shorter-wavelength UV spectral slope $\beta_{18}$ and discuss its utility as an extinction diagnostic in both nearby and high-redshift star-forming galaxies (Section 5). Finally, we discuss our conclusions and plans for future work (Section 6).

\section{Observations}

We originally selected a sample of 31 nearby ($0.003<z<0.029$) star-forming blue compact galaxies (BCGs) with optical spectra from a larger set of 97 galaxies presented in \citet{Kong02.2} and \citet{Kong02.1}. Rest-frame optical spectra of the galaxies' bright nuclei were obtained with the OMR spectrograph on the 2.16m telescope of the Beijing Astronomical Observatory, using a 2\arcsec--3\arcsec slit width that was adjusted based on seeing conditions and, where possible, aligned with the parallactic angle. \citet{Levesque10.1} then selected 36 of these BCGs (restricting the sample to star-forming galaxies based on the criteria of \citealt{Kewley06.1}), and used optical emission line diagnostics to determine a number of ISM properties. These include two different metallicity diagnostics: the $R_{23}$ calibration of \citet{Kobulnicky04.1} and the {\it O3N2} calibration of \citet{Pettini04.1}; for more discussion see Section 4.1. Subsequent study revealed that 5 of the \citet{Levesque10.1} BCGs are actually pairs of galaxies; these were excluded from our final sample.

17 of these star-forming galaxies have been successfully observed by the Cosmic Origins Spectrograph (COS) onboard the Hubble Space Telescope (HST) as part of SNAP program \#13481 (PI: Levesque).  We used the low-resolution G140L grating at the 1105\AA\ central wavelength setting to provide continuous spectral coverage over the range $\sim1150-2200$ \AA\ at a spectral resolution of $R\approx1800$ ($\Delta v\sim200\rm km~s^{-1}$). Each observation consisted of four exposures at different focal plane (FP) positions settings to dither instrumental features over the detector and increase overall signal to noise. Total exposure time per target varied from less than a minute to several kiloseconds; exposure times for each observed galaxy in our sample are given in Table 1. As of this writing the completion rate of this SNAP program is 55\%; one galaxy (2MASX J05554264+0323322) was observed with a truncated exposure time of 87 seconds due to drift from single star guiding effects, and two additional galaxies (III Zw 43 and Haro 3) also had insufficient S/N for our analyses; these data were not included in this work. Our final sample of 14 galaxies, including COS exposure times, is given in Table 1.

The exposures were obtained from the Mikulski Archive for Space Telescopes (MAST). The exposures were binned by 3 pixels ($\sim1/2$ of a resolution element) to improve S/N prior to cross-correlation and coaddition. The calibrated, one-dimensional spectra for each target were next aligned and coadded using the custom IDL procedures described in detail by \citet{Danforth10.1}\footnote{IDL routines available at {\tt \url{http://casa.colorado.edu/$\sim$danforth/costools.html}}}. Briefly, each exposure of a given target was aligned by cross-correlating on a strong interstellar absorption feature.  The data were then interpolated to a common wavelength scale and the flux and error at each wavelength were combined using an exposure-time weighted algorithm. Pixels with instrumental features were de-weighted or excluded from the final coaddition. For our analyses the reduced spectra were adjusted to rest-frame wavelengths (redshifts are given in Table 1).

\begin{figure}[!htbp]
\epsscale{0.32}
		\plotone{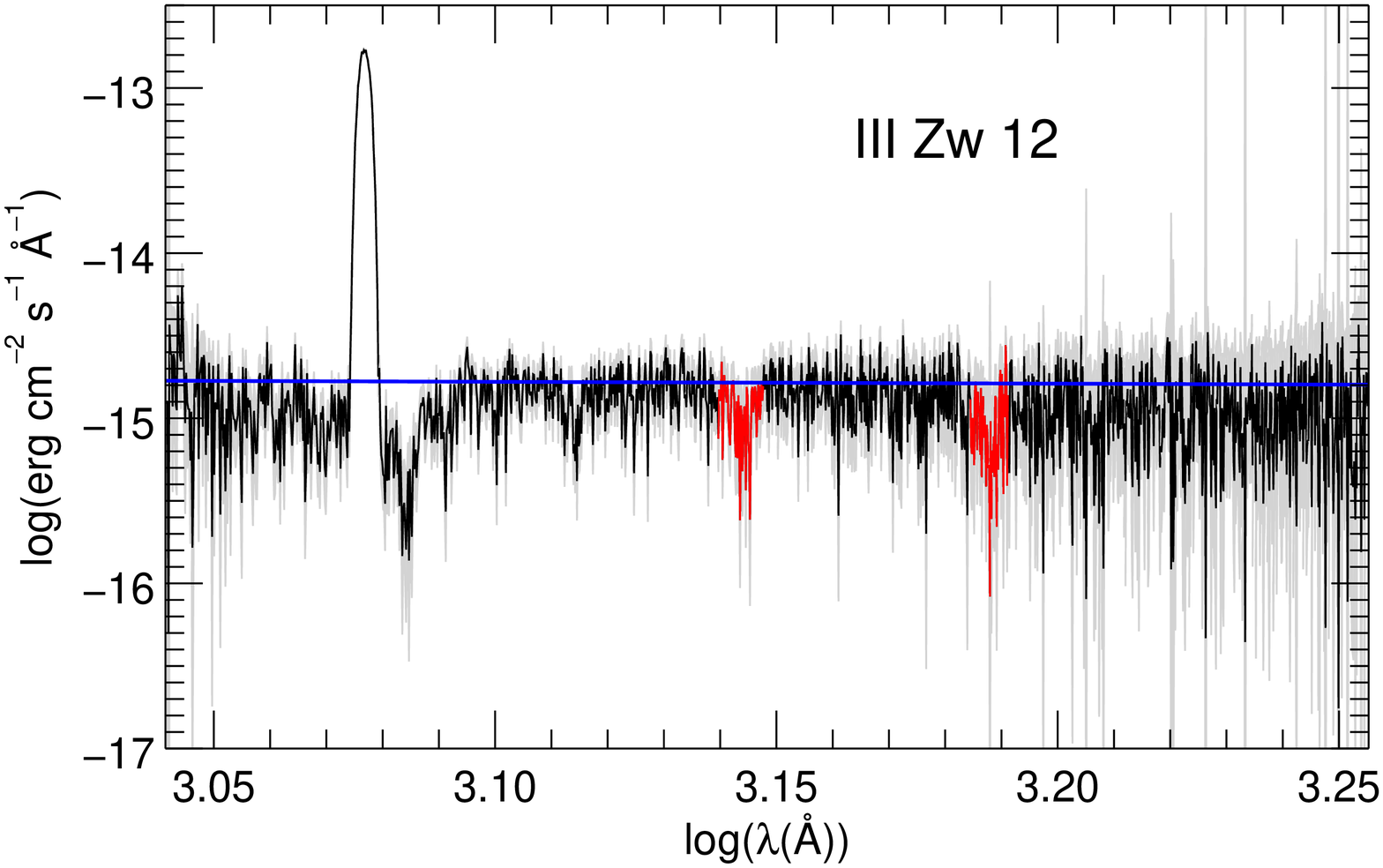}
		\plotone{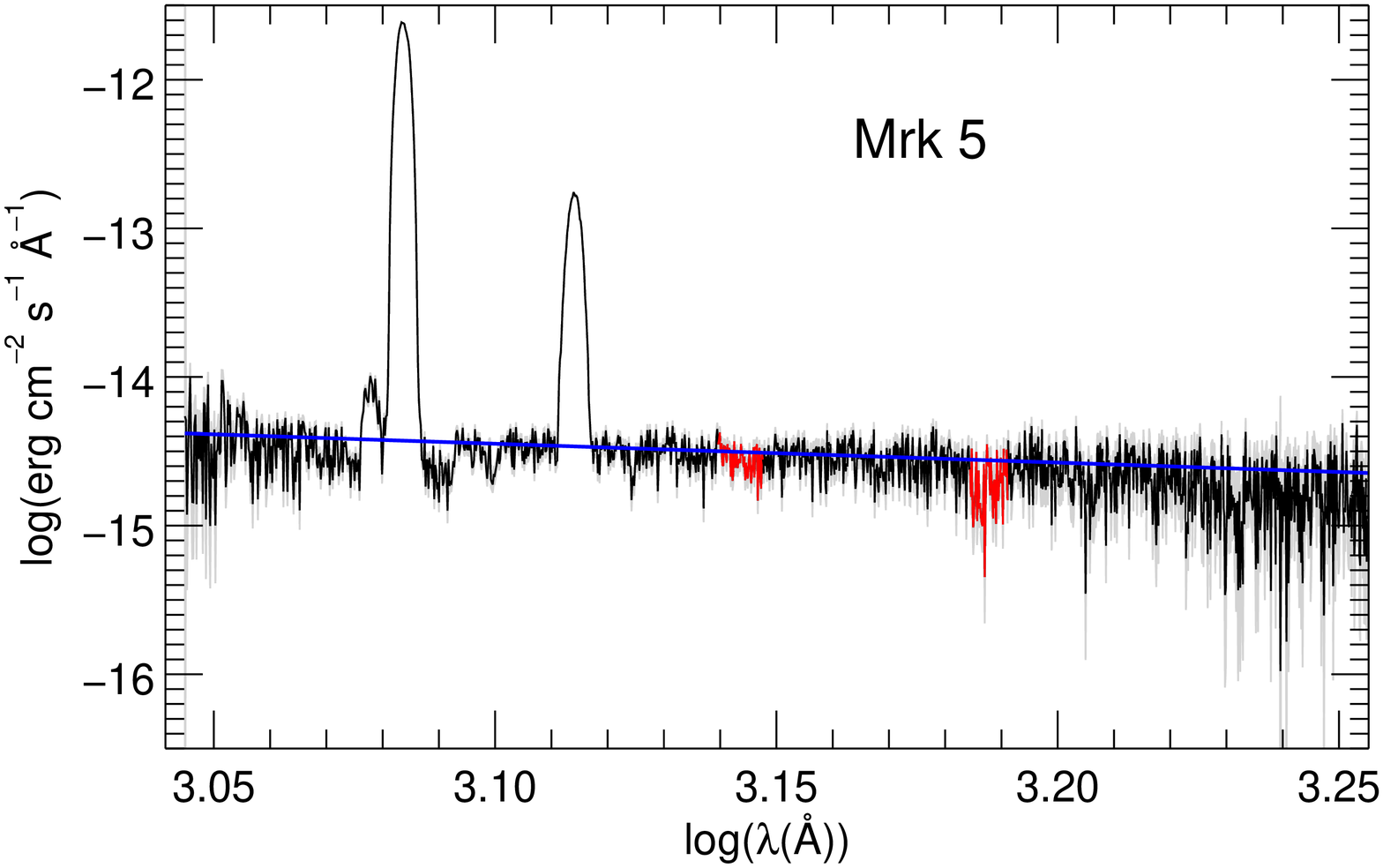}
		\plotone{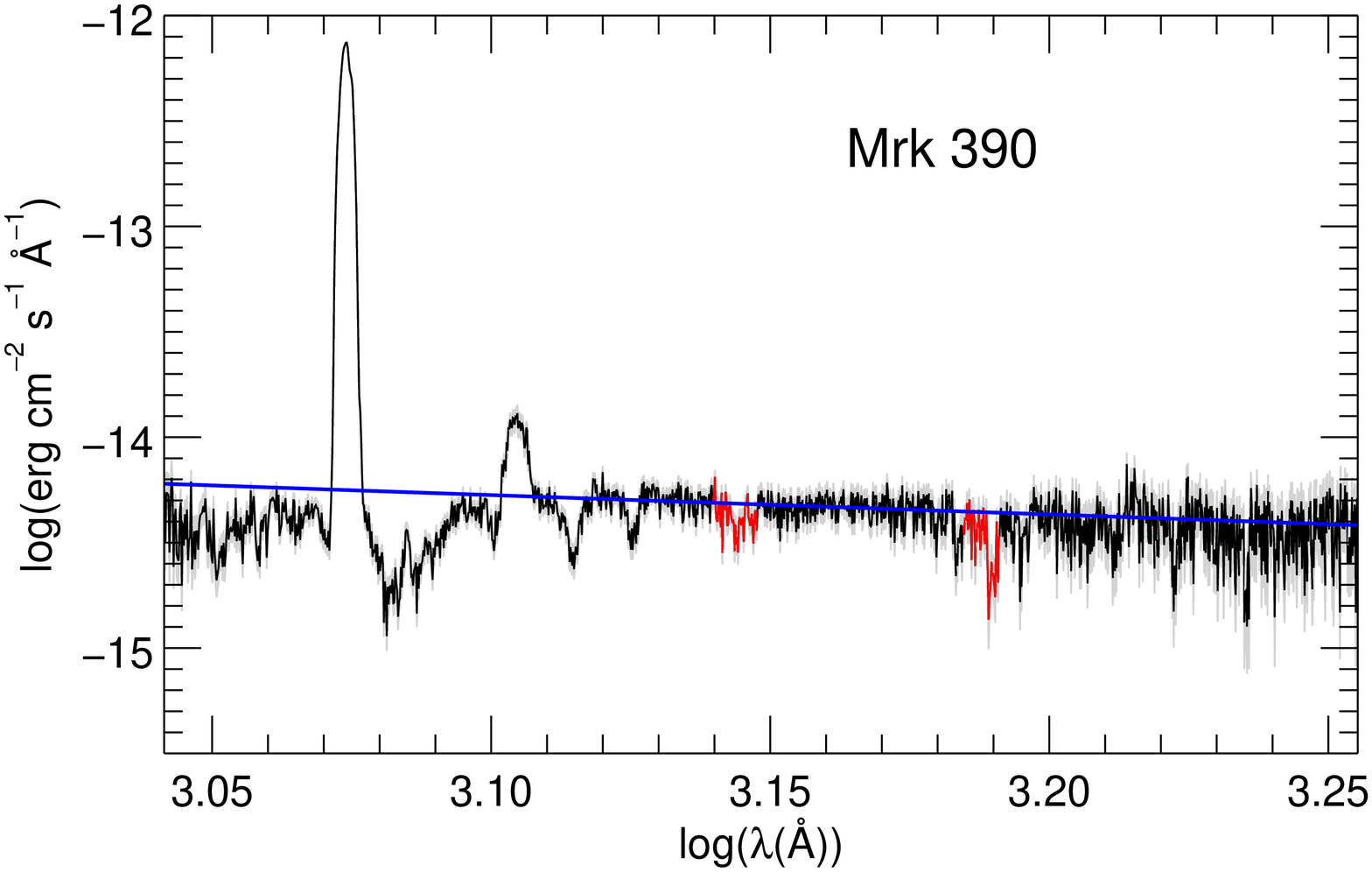}
		\plotone{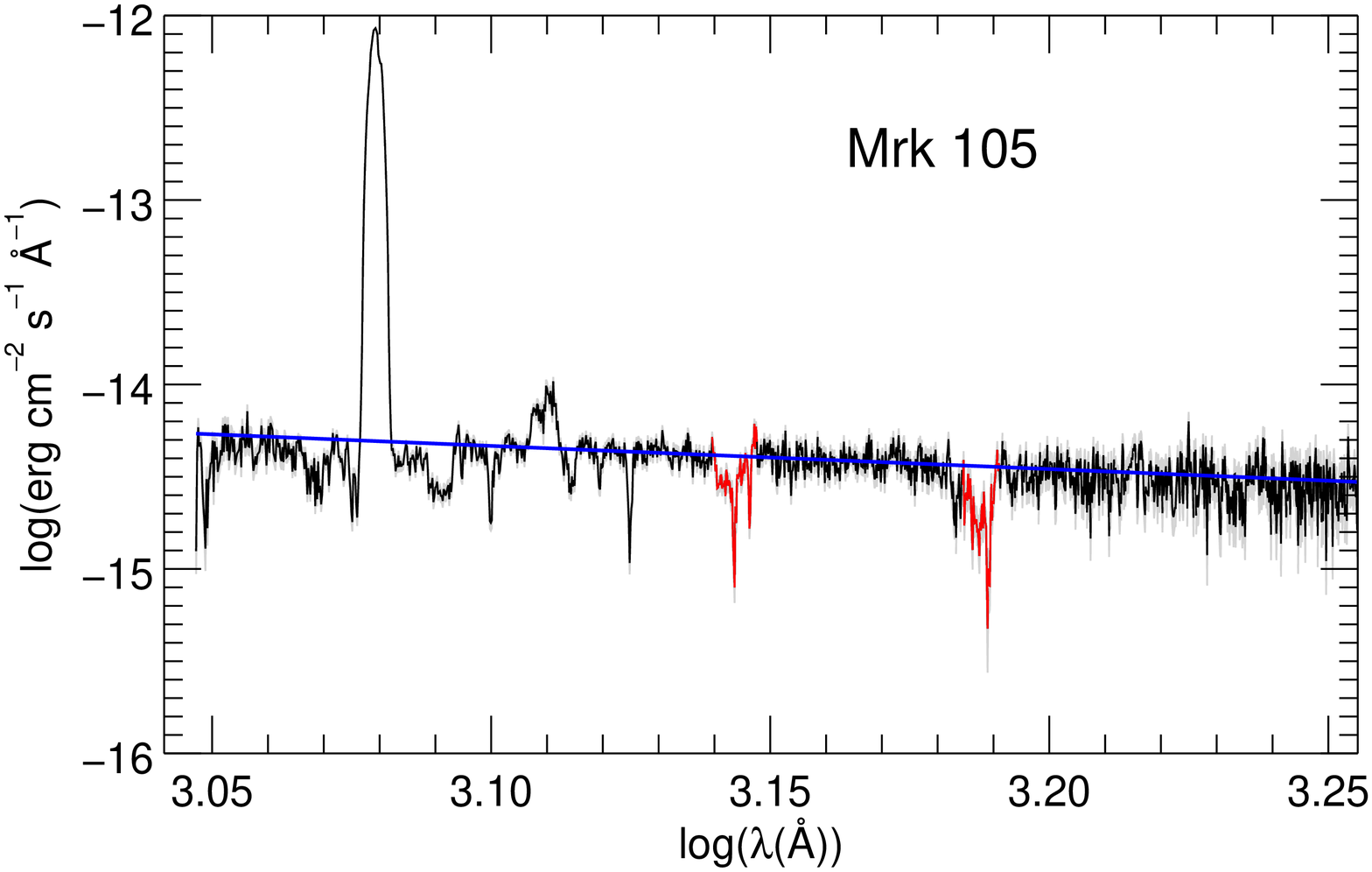}
		\plotone{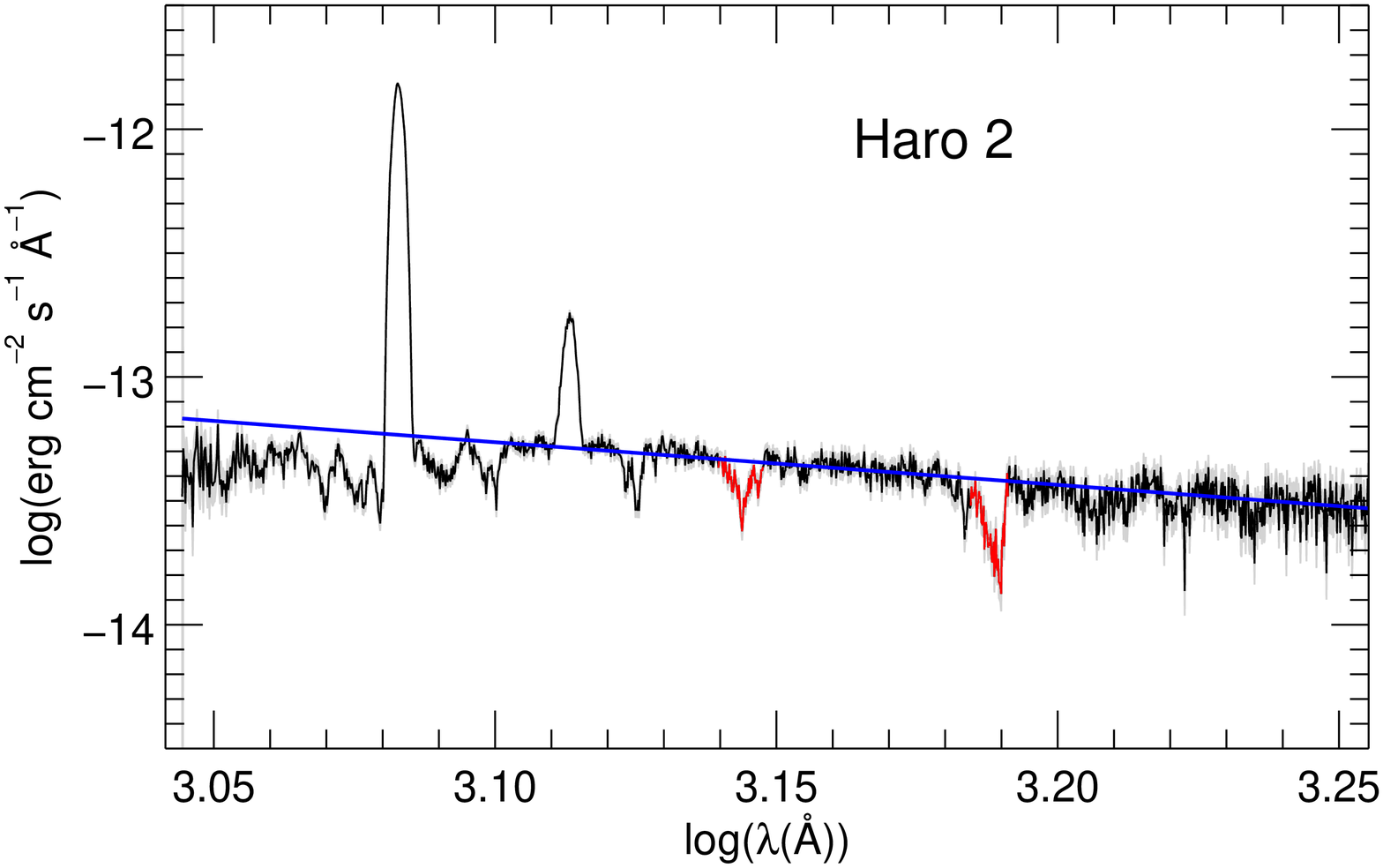}
		\plotone{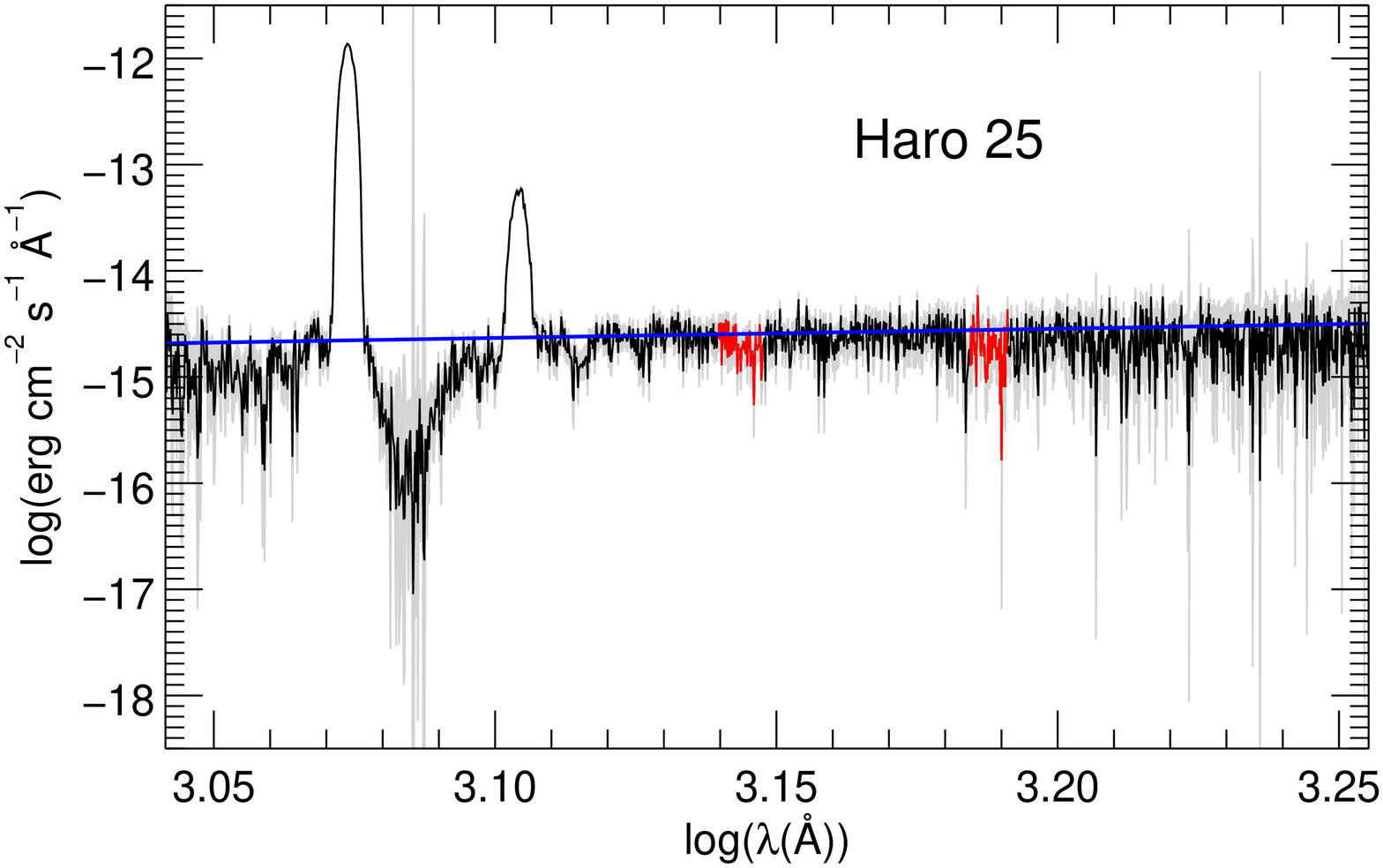}
		\plotone{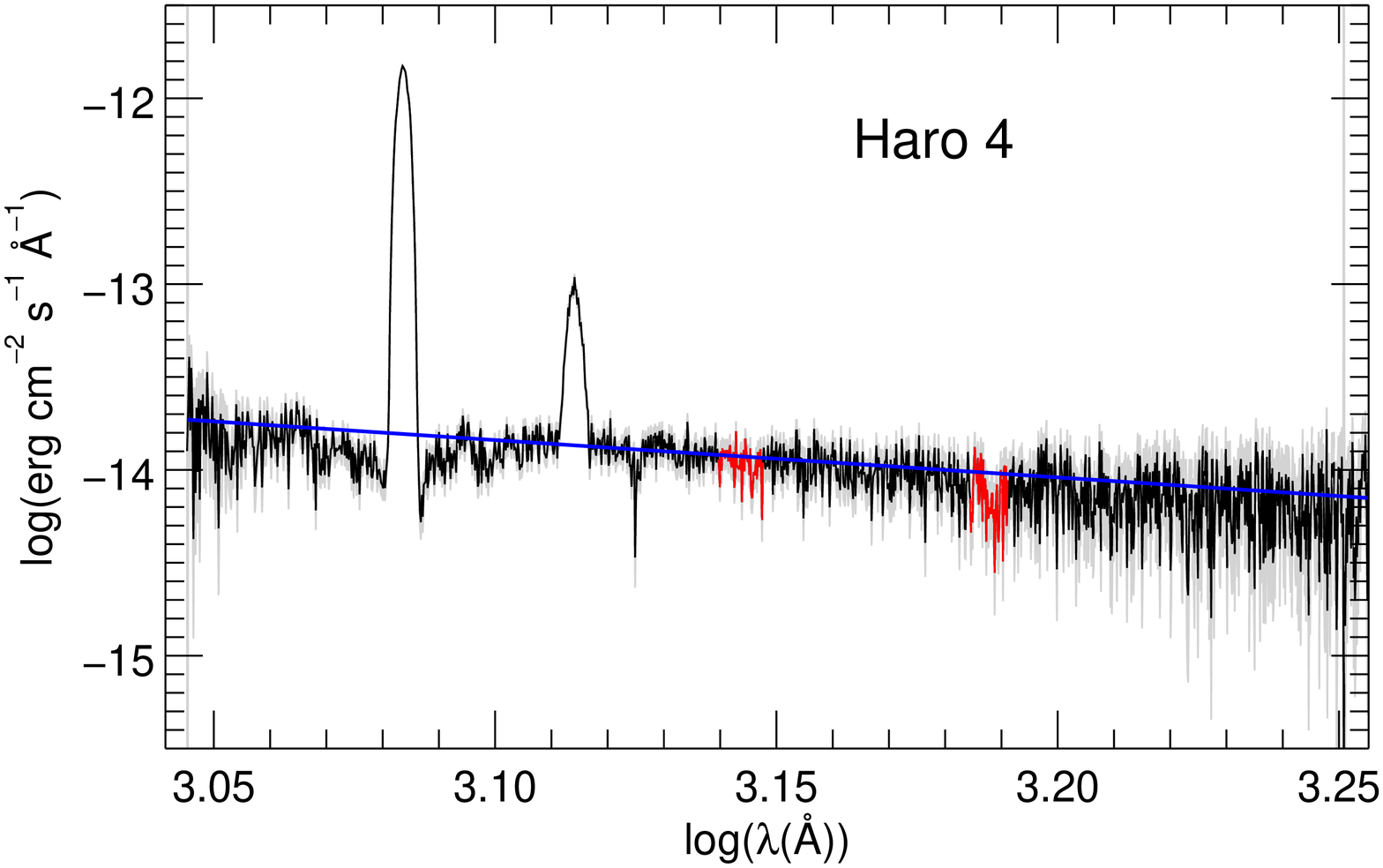}
		\plotone{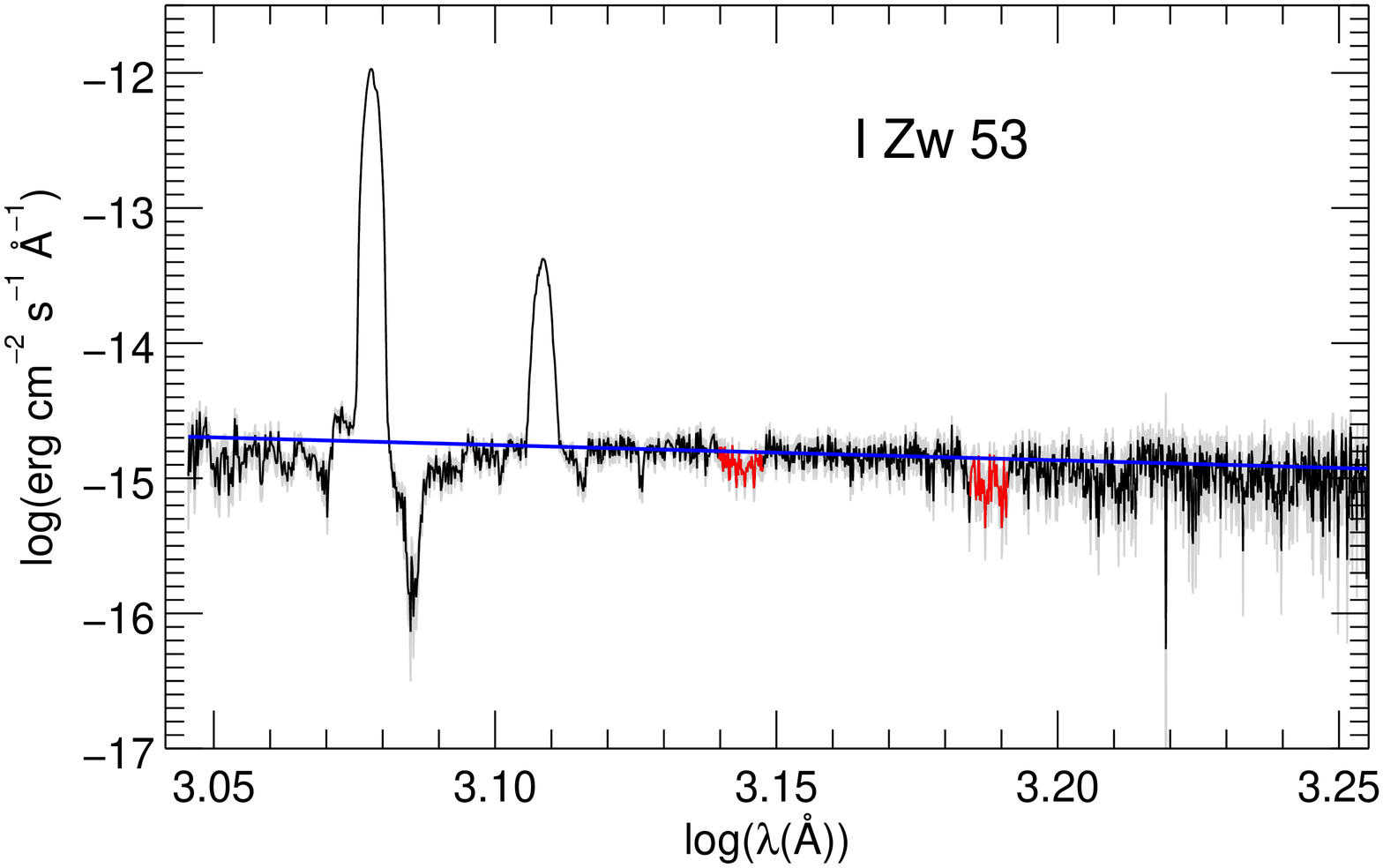}
		\plotone{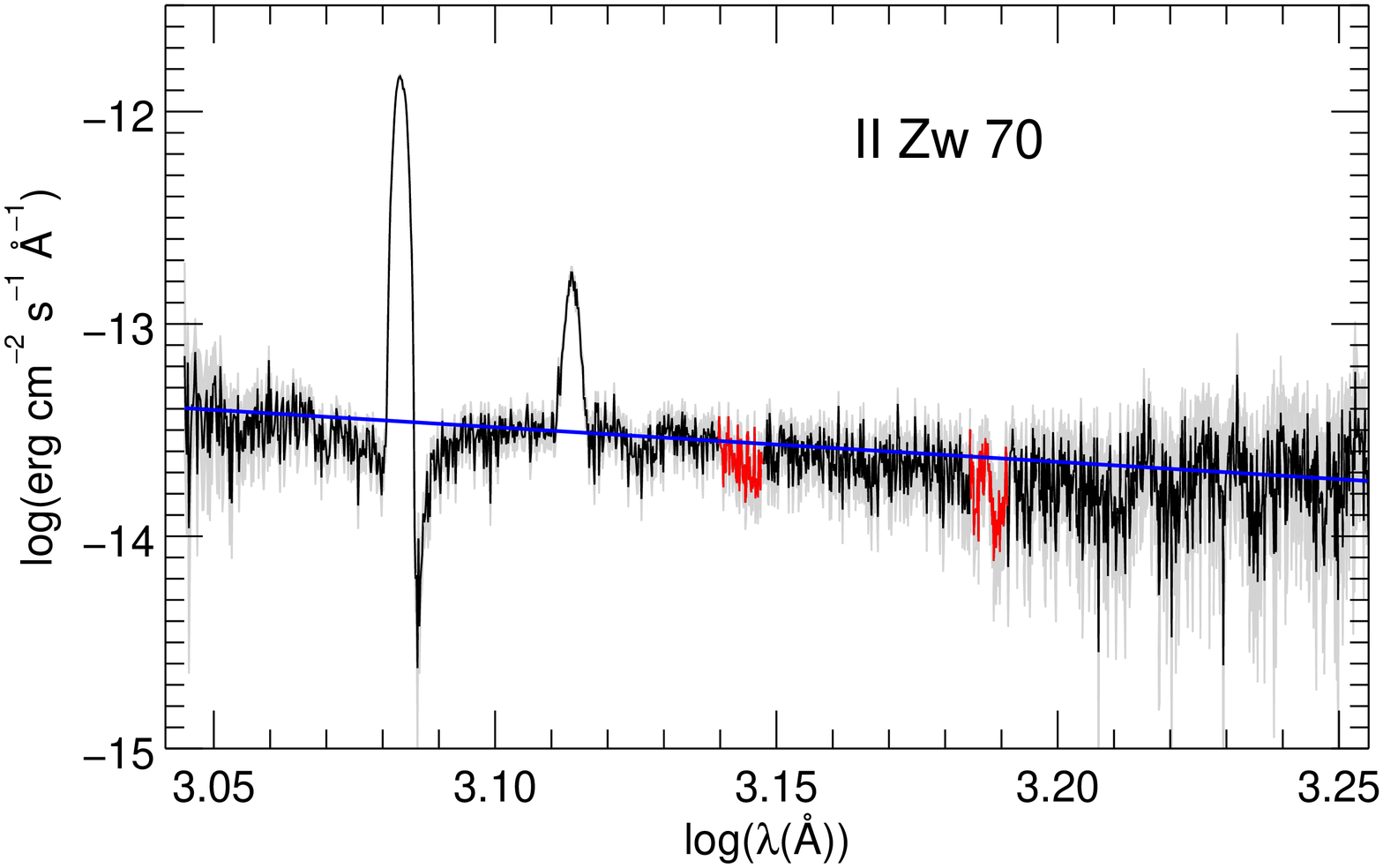}
		\plotone{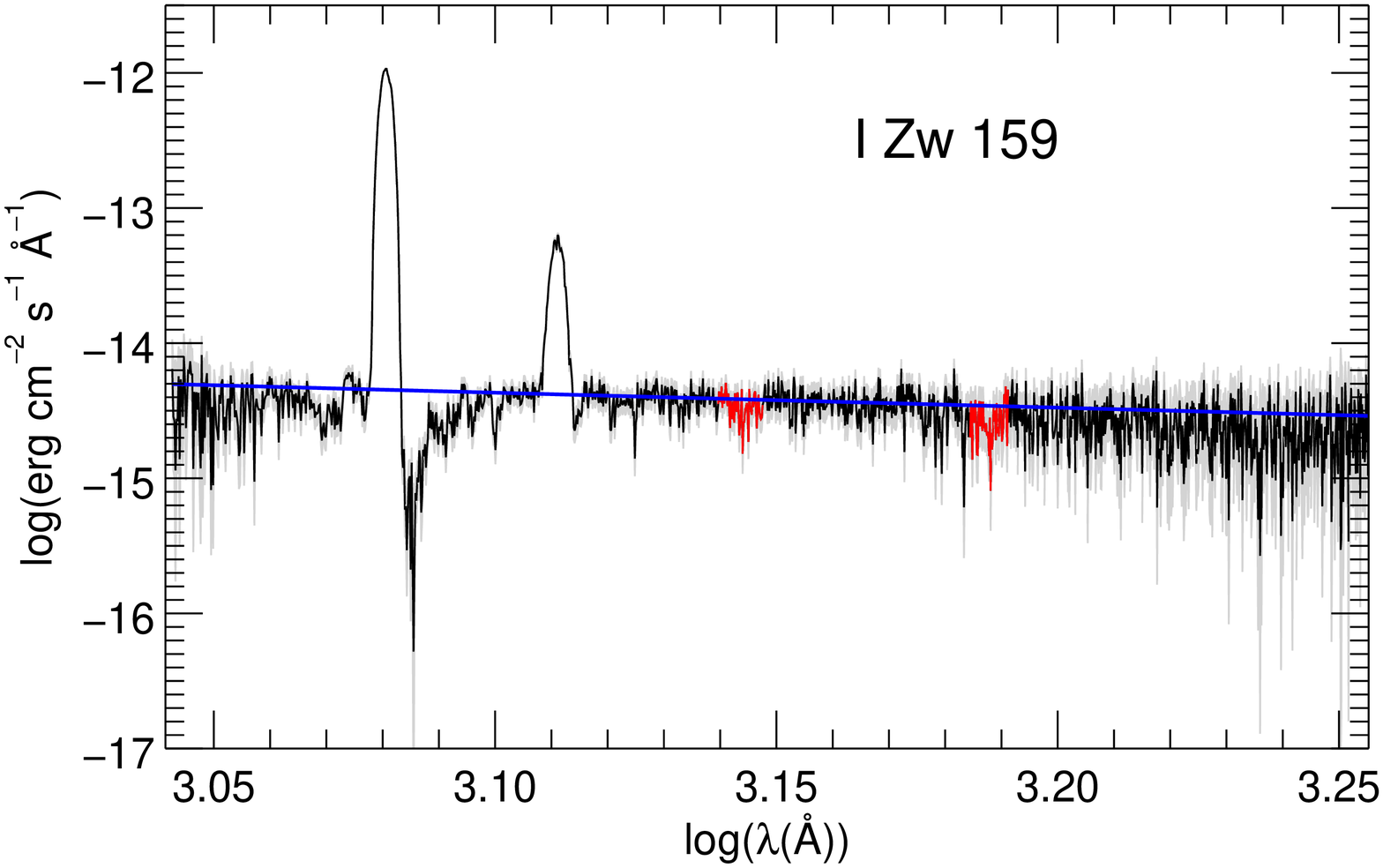}
		\plotone{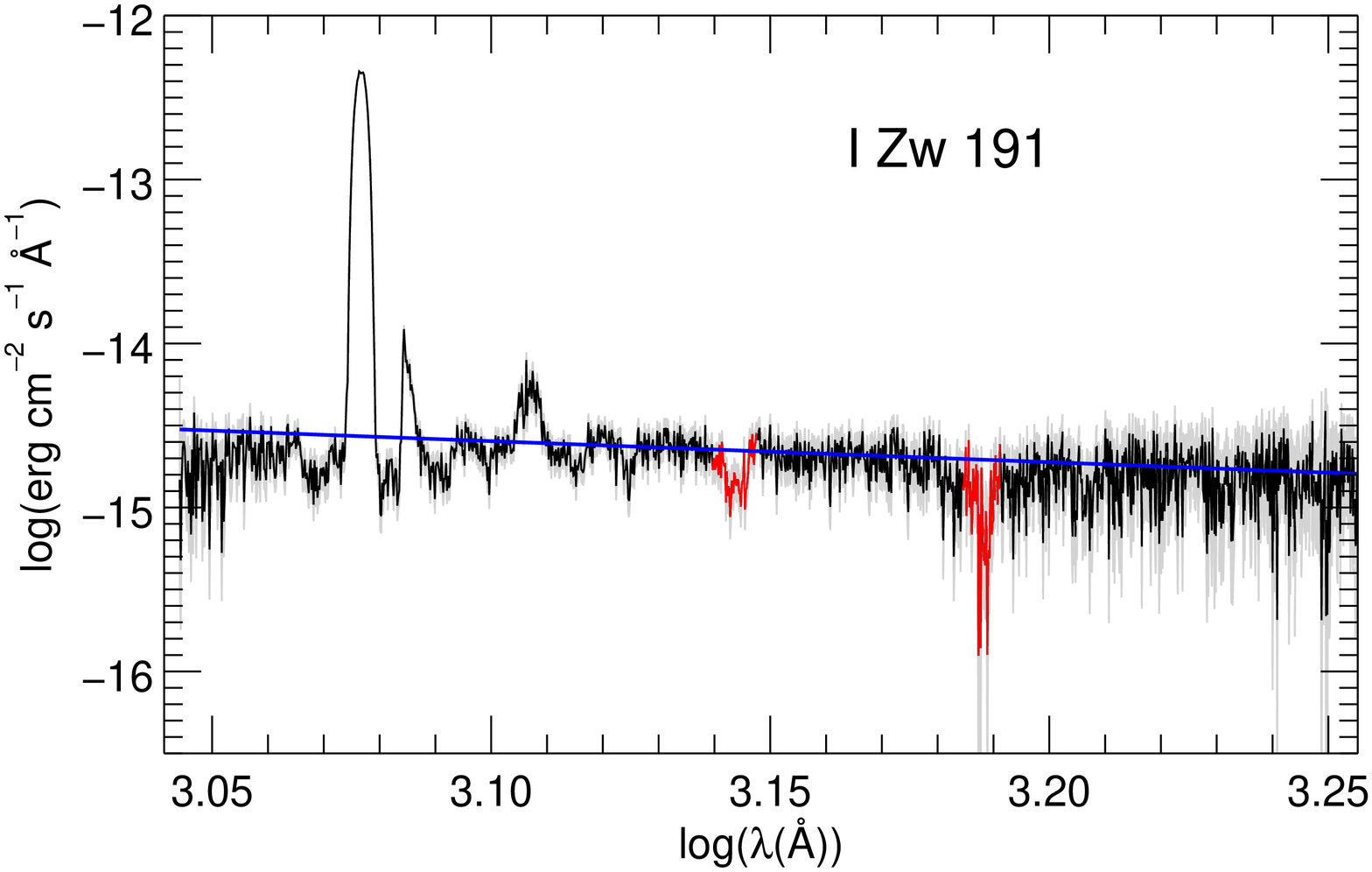}
		\plotone{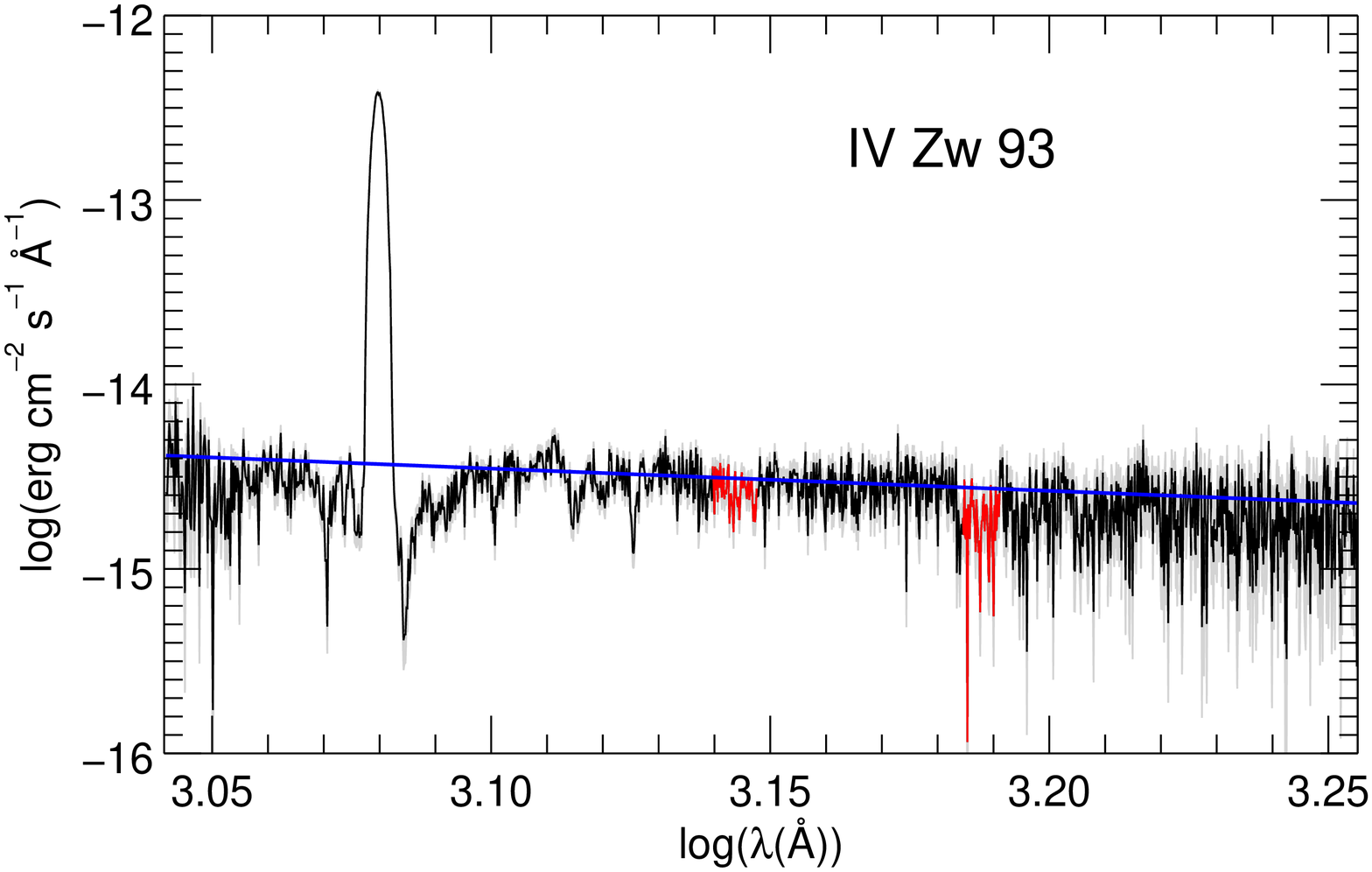}
		\plotone{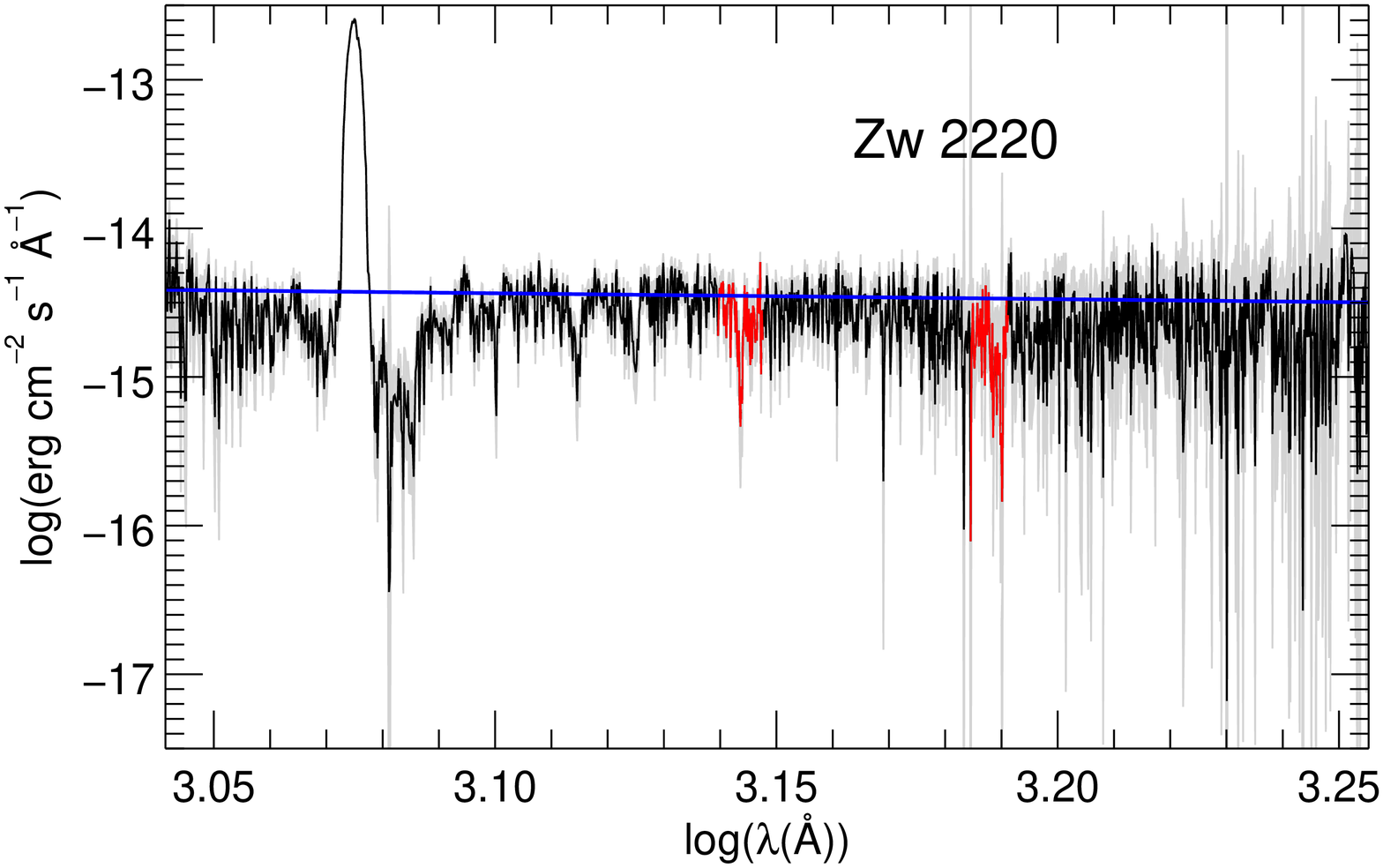}
		\plotone{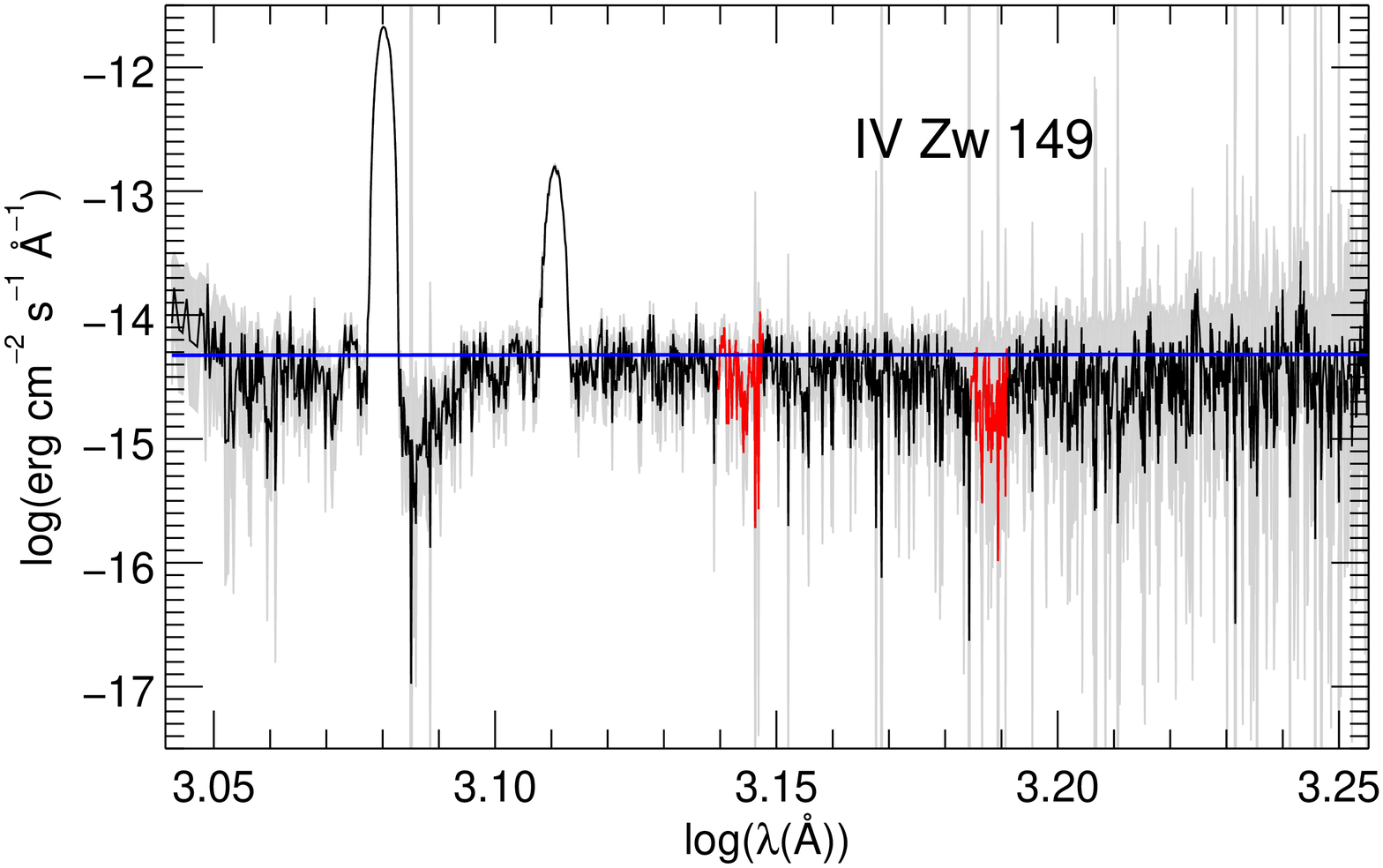}
	\caption{HST COS spectra of our star-forming galaxies. Spectra have been corrected to rest-frame; the strong emission features near $\lesssim$1215\AA\ and $\lesssim$1302\AA\ represent blueshifted geocoronal Ly$\alpha$ and O\textsc{i}, respectively. Red regions of the spectra illustrate the wavelengths of the Leitherer SiIV\_1400 and CIV\_1550 diagnostic indices; see Section 4.2 for further discussion. The spectra are plotted in log space to best illustrate our fits (blue line) of the UV spectral slope, $\beta_{18}$ (where f($\lambda$) $\propto \lambda^{\beta_{18}}$ for $1250 < \lambda < 1800$ \AA; see Section 5). Variations as determined from adding and subtracting the error spectra are shown in gray.}
	\label{fig:ObsSpectra}
\end{figure}

\begin{deluxetable}{l c c c c c c c c c}
\tabletypesize{\scriptsize}
\tablewidth{0pc}
\tablecaption{Blue Compact Galaxies Observed with HST}
\tablecolumns{10}
\label{tab:TargetList}
\tablehead{
\colhead{Galaxy}
& \colhead{RA}
& \colhead{Dec}
& \colhead{$z$}
& \colhead{$B$}
& \multicolumn{3}{c}{log(O/H)+12\tablenotemark{a}}
& \colhead{SFR\tablenotemark{a}}
& \colhead{Exp. Time} \\ \cline{6-8}
\multicolumn{5}{c}{}
&\colhead{KK04}
&\colhead{O3N2}
&\colhead{N2}
&\colhead{(M$_{\odot}$ yr$^{-1}$)}
& \colhead{(s)}
}
\startdata
III Zw 12 & 00 47 56.3 & +22 22 22.5  & 0.019  &15.0  &8.57 & 8.52  &8.51            &  1.58 & 720\\ 
Mrk-5 &06 42 15.9 & +75 37 40.0 & 0.003  &17.0  &8.64 & 8.11    &8.13              &  0.03 & 1080 \\ 
Mrk-390 &08 35 33.1 & +30 32 03.2 & 0.025   &15.0 &8.41 & 8.42   &8.42          &  9.31 &1568 \\
Mrk-105 &09 20 26.3 & +71 24 16.0 & 0.013  &16.0  &8.94 & 8.56  &8.58              &  0.86 &2060  \\
Haro 2 & 10 32 32.0 & +54 24 03.5 & 0.005 &13.7   &8.72 & 8.41  &8.39               &  1.03 &420 \\
Haro 25 & 10 48 44.2 & +26 03 13.2 & 0.026 &15.0 &8.76 & 8.33  &8.36             &  11.08 &568 \\ 
Haro 4 & 11 04 58.5 & +29 08 22.1 & 0.003   &15.5 &8.22 & 7.97   &7.96              &  0.02 &260 \\
I Zw 53 &13 13 57.7 & +35 18 52.8  & 0.016  &16.8  &8.66 & 8.55   &8.62             &  0.56  &2200 \\
II Zw 70  &14 50 56.5 & +35 34 18.2  & 0.004  &14.5  &8.09 & 8.14  &8.09              &  0.08 &80 \\
I Zw 159 &16 35 21.1 & +52 12 52.8  & 0.010 &15.6   &8.56 & 8.25  &8.26               &  0.52  &620  \\
I Zw 191 &17 40 24.8 & +47 43 58.6 & 0.019 &14.8   &8.92 & 8.73   &8.72          &  3.23  &1120  \\
IV Zw 93 &22 16 07.7 & +22 56 32.7 & 0.012 &15.7   &8.40 & 8.26  &8.22              &  0.57  &1120  \\ 
Zw 2220 &22 23 00.6 & +30 55 23.0 & 0.023 & 14.0   &8.88 & 8.59  &8.71           &  4.90  &380  \\
IV Zw 149 &23 27 41.1 &  +23 35 20.2 & 0.011 &13.2   &8.47 & 8.41  &8.49                &  0.73  &140  \\
\enddata
\tablenotetext{a} {Based on analyses in \citet{Levesque10.1}, adopting raw emission line fluxes from \citet{Kong02.1} and correcting for the total line-of-sight extinction.}
\end{deluxetable}

The fully-combined data vary wildly in terms of data quality due mostly to source flux and exposure time. Since many of the star-forming galaxies in our sample are extended objects which fill the COS aperture, resolution for some targets may be degraded from the nominal $R\approx1800$ typical of point sources observed with COS/G140L; however, BCGs are dominated by their central clusters and thus appear close to point-like, yielding only minimal loss of resolution in COS observations. It should also be noted that the 2.5\arcsec diameter field stop of the COS Primary Science Aperture is comparable to the 2\arcsec--3\arcsec slit widths used in the \citet{Kong02.1} observations; by centering COS on the bright nuclei of the BCGs we sample comparable galaxy regions with our FUV spectra to those observed with the \citet{Kong02.1} optical spectra. Our spectra are shown in Figure 1.

\section{Stellar Population Synthesis Models}
Starburst99 \citep{Leitherer99.1,Vazquez05.1,Leitherer09.1,Leitherer14.1} is a stellar population synthesis code which produces model spectral energy distributions as a function of metallicity, star formation, initial mass function (IMF), age, and evolution of the simulated stellar populations. This is done by combining model stellar atmospheres and spectra with grids of stellar evolutionary tracks.

For this work we used this code to generate a grid of model ionizing spectra with varied stellar evolutionary models, metallicities, IMFs, and treatments of the star formation history. We used three different sets of evolution tracks produced by the Geneva group. The tracks from \citet{Meynet94.1} (hereafter M94) simulate the evolution of a non-rotating stellar population with enhanced mass-loss rates. We also adopt the rotating and non-rotating tracks of \citet{Ekstrom12.1} and \citet{Georgy13.1}; the rotating tracks (hereafter ROT), include the effects of stellar rotation with initial rotation rates of $v = 0.4$ of the break-up velocity, which is the peak of the velocity distribution in young B stars \citep{Huang10.1}, while the non-rotating tracks (hereafter NOROT) simulate identical initial conditions but do not include rotation.

The metallicities of our models were based on the assumed abundances for the evolution tracks. The M94 tracks span five fractional heavy element abundances by mass ($Z = $ 0.001, 0.004, 0.008, 0.020, and 0.04), while the ROT and NOROT tracks are available at $Z = 0.002$ and $Z = 0.014$. The abundance for the optical high resolution spectrum in Starburst99 is set separately, and was chosen for each simulation to most closely match the abundance of the chosen stellar evolutionary tracks.

We initially adopted a Kroupa IMF \citep{Kroupa01.1}, a power-law in $dN/dM$ with $\alpha = 1.3$ for the $0.1 - 0.5$ M$_\odot$ mass range, and $\alpha = 2.3$ for the $0.5 - 100$ M$_\odot$ mass range. In addition, top-heavy and flattened IMF models were made with high-mass slopes set to $\alpha = 1.3$ and $\alpha = 3.3$ while keeping the low-mass slope set at $\alpha = 1.3$. As massive O and B stars are the main contributors to the UV spectrum, we are primarily concerned with the high-mass slope. Unless otherwise stated, all models presented here assume a Kroupa IMF.

Our models simulated both an instantaneous burst of star formation and a region of continuous star formation. In models with an instantaneous burst of star formation, a zero-age stellar population of $10^6$ M$_\odot$ is simulated, allowing us to trace the effects of a single coeval stellar population; we modeled the evolution of the stellar population up to 20 Myr in 0.5 Myr increments. The continuous star formation models adopt a constant rate of 1 M$_\odot$ yr$^{-1}$, representing the other extreme of star formation history. Here, while we also model the evolution of the population in 0.5 Myr increments, we consider only the models at an age of 5 Myr, the age at which a continuously star-forming stellar population reaches equilibrium (at younger ages this equilibrium is not yet reached; at older ages there is little to not evolution of the stellar FUV ionizing spectrum. For more discussion see \citealt{Kewley01.1}).

\section{Metallicity Diagnostics}

\subsection{Optical}

Using the existing optical spectra for our observed galaxies from \citet{Kong02.2} and \citet{Kong02.1} we can calculate the metallicities (defined here as the oxygen abundance 12 + log(O/H)) using several common strong-line optical diagnostics.

\citet{Kobulnicky04.1} (hereafter KK04) used the stellar evolution and photoionization grids of \citet{Kewley02.1} to derive their metallicity diagnostic. Like many optical metallicity diagnostics, this method depends on the ([O \textsc{ii}] $\lambda$3727 + [O \textsc{iii}] $\lambda \lambda$4959, 5007)/H$\beta$ line ratio, known as $R_{23}$. Unfortunately $R_{23}$ is double valued with metallicity; KK04 uses the additional line ratio [N \textsc{ii}] $\lambda$6583/[O \textsc{ii}] $\lambda$3727 to break this degeneracy and determine whether to use the upper or lower metallicity result. The KK04 diagnostic works by iterating between calculating ionization parameter and metallicity until the values converge. The ionization parameter formula depends on metallicity and the [O \textsc{iii}] $\lambda \lambda$4959, 5007/[O \textsc{ii}] $\lambda$3727 line ratio. The metallicity formulae depend on the ionization parameter and $R_{23}$, and have upper and lower branch versions. If the lower branch gives a higher metallicity than the upper branch, the metallicity is set to 12 + log(O/H) = 8.4, the cutoff between the two branches. The estimated accuracy of this diagnostic is $\sim 0.15$ dex.

\citet{Pettini04.1} developed two optical metallicity diagnostics using $T_e$-based metallicities and observations of H \textsc{ii} regions. The first of these diagnostics (hereafter O3N2) is a linear fit which depends on the line ratio log[(O \textsc{iii}] $\lambda$5007/H$\beta$)/([N \textsc{ii}] $\lambda$6583/H$\alpha$)]. It is valid for log[(O \textsc{iii}]/H$\beta$)/([N \textsc{ii}]/H$\alpha$)] $< 2$. The estimated accuracy of the O3N2 diagnostic is $\sim 0.14$ dex. The second diagnostic (hereafter N2) is a cubic fit which depends on the line ratio [N \textsc{ii}] $\lambda$6583/H$\alpha$. It is valid for $-2.5 <$ [N \textsc{ii}]/H$\alpha$ $< -0.3$. The estimated accuracy of the N2 diagnostic is $\sim 0.18$ dex. For a more detailed discussion of optical metallicity diagnostic development, see \citet{Kewley08.1}.

\subsection{UV - Leitherer Indices}

\citet{Leitherer11.1} established a set of UV indices for star-forming galaxy spectra. Each index is a blend of lines, mostly of interstellar or stellar wind origin. Three wavelength regions are defined for each index: a blue continuum, a central index bandpass, and a red continuum. Within each continuum range the median flux in calculated, as well as the midpoint in the wavelength range. The continuum is defined as the line connecting the representative blue continuum and red continuum points. The equivalent width is calculated as the sum of the difference between the data and the continuum in each bin within the central bandpass, normalized by the continuum. The error was determined by calculating the standard deviation of the blue and red continua data from the continuum line. This value was multiplied by the square root of the number of bins in the central index bandpass to find the error of the equivalent width of the index.

Of Leitherer's indices, SiIV\_1400 and CIV\_1550 are the only two dominated by stellar wind lines. The main feature included in SiIV\_1400 is the doublet Si \textsc{iv} $\lambda 1393,1402$, while the main feature included in CIV\_1550 is the doublet C \textsc{iv} $\lambda 1550,1548$.  Both the Si \textsc{iv} $\lambda$1400 doublet and the C \textsc{iv} $\lambda$1550 doublet have interstellar components, which sometimes provide a significant contribution to the line equivalent widths, in addition to their stellar wind components \citep{Leitherer11.1}. \citet{Heckman98.1} finds a correlation with dust extinction and reddening of the vacuum-UV continuum for these high-ionization lines, but a weaker correlation that those found for low-ionization lines.

In their spectra, \citet{Leitherer11.1} found that, by visual inspection, the interstellar component in the Si \textsc{iv} doublet could account for $\sim 2/3$ of the equivalent width, whereas the interstellar component in the C \textsc{iv} doublet could account for $\sim 1/3$ of the equivalent width. While the stellar wind component in the Si \textsc{iv} doublet only becomes apparent for blue giants and supergiant stars \citep{Shapley03.1}, wind effects can be seen in the C \textsc{iv} doublet of early-O main-sequence stars \citep{Leitherer01.1}.

\subsection{Comparison of Optical and UV Diagnostics}

We measured the SiIV\_1400 and CIV\_1550 indices in all of our HST COS spectra (see Table~\ref{tab:COSdata}); the wavelength ranges of the central bandpasses for both indices in each of our galaxies are illustrated in Figure 1. For comparison purposes the COS data were binned by a factor of 2 to approximate the 0.4\AA\ resolution of the Starburst99 spectral output. We also used the optical emission line fluxes given in \citet{Kong02.1} to determine optical strong line metallicities using the KK04, O3N2, and N2 diagnostics for each of our galaxies based on the optical emission line fluxes given in \citet{Kong02.1} and the analyses detailed in \citet{Levesque10.1}.

\begin{deluxetable}{l c c c c}
\tabletypesize{\scriptsize}
\tablewidth{0pc}
\tablecaption{Measured Spectroscopic Properties of BCGs Observed with HST}
\tablecolumns{6}
\label{tab:COSdata}
\tablehead{
\colhead{Galaxy}
& \colhead{SiIV\_1400}
& \colhead{C IV\_1550}
& \colhead{$\beta_{18}$\tablenotemark{a}}
& \colhead{$\tau^l_B$}\tablenotemark{a,b}
\\
\colhead{}
& \colhead{(\AA)}
& \colhead{(\AA)}
& \multicolumn{2}{c}{}
}

\startdata
III Zw 12   & $5.41 \pm 2.74$ & $8.56 \pm 2.98$ & $-0.12 \pm 0.14$ &$0.25 \pm 0.04$  \\
Mrk 5      & $1.47 \pm 1.64$ & $4.97 \pm 2.10$ & $-1.27 \pm 0.06$ &$-0.04 \pm 0.01$     \\
Mrk 390   & $3.60 \pm 0.77$ & $5.83 \pm 1.14$  & $-0.92 \pm 0.05$ &$0.47 \pm 0.04$     \\
Mrk 105   & $4.74 \pm 0.93$ & $9.31 \pm 1.10$ & $-1.26 \pm 0.04$ &$0.42 \pm 0.04$      \\
Haro 2    & $5.11 \pm 0.48$ & $7.98 \pm 0.67$ & $-1.72 \pm 0.03$ &$0.43 \pm 0.03$     \\
Haro 25   & $3.39 \pm 2.20$ & $4.74 \pm 2.24$   & $0.87 \pm 0.13$  &$0.26 \pm 0.03$      \\
Haro 4    & $1.45 \pm 1.37$ & $3.00 \pm 1.78$   & $-2.01 \pm 0.08$ &$0.01 \pm 0.01$     \\
I Zw 53   & $5.48 \pm 1.23$ & $6.65 \pm 1.42$ & $-1.13 \pm 0.08$ &$0.65 \pm 0.18$     \\
II Zw 70   & $4.47 \pm 1.31$ & $2.70 \pm 1.98$   & $-1.63 \pm 0.10$ &$-0.08 \pm 0.03$      \\
I Zw 159   &$2.34 \pm 1.30$ & $3.63 \pm 2.18$ & $-1.11 \pm 0.09$ &$0.16 \pm 0.01$     \\
I Zw 191   & $5.78 \pm 1.38$ & $9.27 \pm 1.76$  & $-1.28 \pm 0.10$ &$0.57 \pm 0.04$     \\
IV Zw 93  & $2.25 \pm 1.88$ & $6.89 \pm 2.02$ & $-1.21 \pm 0.07$ &$0.22 \pm 0.03$     \\
Zw 2220    & $6.74 \pm 2.37$ & $8.24 \pm 3.18$  & $-0.41 \pm 0.12$ &$0.32 \pm 0.06$     \\
IV Zw 149  & $1.49 \pm 3.11$ & $2.89 \pm 5.72$  & $0.02 \pm 0.21$ &$0.28 \pm 0.01$     \\ 
\enddata
\tablenotetext{a} {Corrected for Galactic foreground reddening based on \citet{Schlegel98.1} and \citet{Schlafly11.1}.}
\tablenotetext{b} {$\tau^l_B =$ ln((H$\alpha$/H$\beta$)/2.86), following the definition of \citet{Calzetti94.1}}
\end{deluxetable}

In Figure 2 we compare the two Leitherer indices to the three optical diagnostic metallicities. There are statistically robust positive linear relations (Spearman's $\rho \ge 0.6$, $p \le 0.02$) between the SiIV\_1400 and CIV\_1550 indices and all three of the optical diagnostic metallicities.

\begin{figure}[!ht]
		\includegraphics[width=\textwidth] {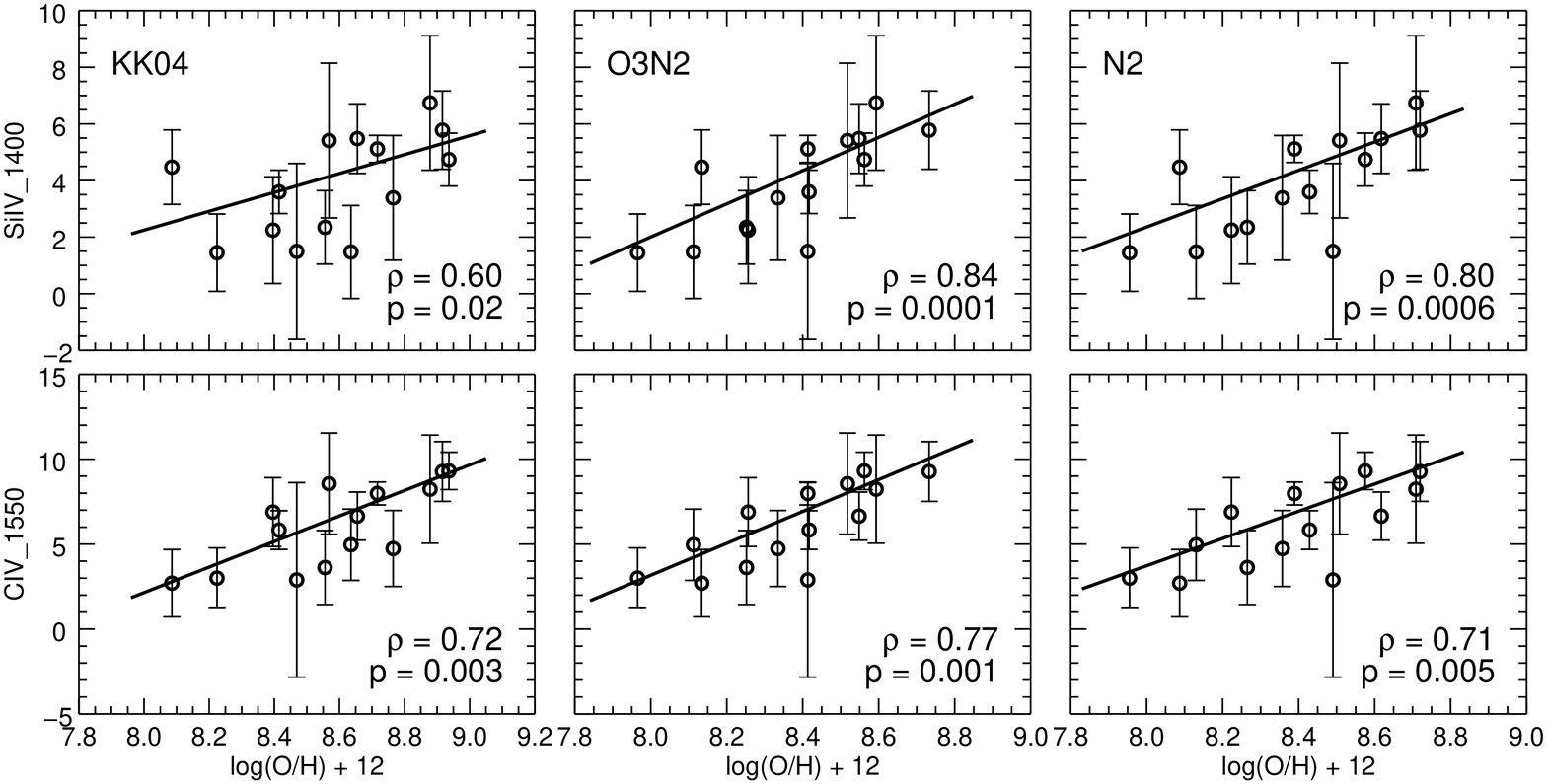}
	\caption{The SiIV\_1400 (top) and CIV\_1550 (bottom) indices measured from our HST spectra and compared to the KK04 (left), O3N2 (center), and N2 (right) metallicities determined from the optical data of \citet{Kong02.1}. Solid lines illustrate the best linear fits, and Spearman's $\rho$ and p-values are given for each pair of diagnostics.}
	\label{fig:Leitherer_HST}
\end{figure}

We measured the SiIV\_1400 and CIV\_1550 Leitherer indices in our full grid of Starburst99 models using the same definitions and calculation procedures detailed in Section 4.2. The indices were measured from the non-normalized \texttt{.ifaspec} Starburst99 output spectra; these are theoretical UV spectra calculated between 900--3000\AA\ at a resolution of $\sim$0.4\AA, adopting WM-Basic models for the O star atmospheres \citep{Pauldrach01.1} and the Potsdam models for Wolf-Rayet atmospheres \citep{Hamann04.1}; for a detailed discussion see \citet{Leitherer10.1}.

It should be noted that, as an evolutionary synthesis code, Starburst99 alone does not include any treatment of photoionization or contributions from the ISM. As a result, the interstellar contributions to the Leitherer indices seen in observations are not present; the models only simulate the stellar components (wind and photospheric) of these indices.

The index definitions of \citet{Leitherer11.1} also include an inherent metallicity bias. The definitions of the indices' continuua are themselves metallicity dependent and thus measure a ``continuum" level that is lower at high metallicities due to line blanketing effects, leading to a slight decrease in the equivalent widths of indices measured at high metallicities. A positive relation between index value and metallicity is therefore slightly ``flatter" under this definition than it would be for a metallicity-corrected continuum definition (with the opposite holding through for a negative relation). For this reason we have measured the index values in our Starburst99 models from the non-normalized synthetic spectra, retaining the effects of a metallicity-dependent index continuum. These synthetic spectra are fully blanketed and are not adjusted for knowledge of the true continuum. Thus the same bias is present in both our observed and theoretical spectra.

The evolution of the SiIV\_1400 and CIV\_1550 indices from our Starburst99 models as a function of metallicity are shown in Figures 3 and 4. Figure 3 shows models adopting an instantaneous burst star formation history at 1, 3, 5, and 7 Myr, while Figure 4 shows models adopting a continuous star formation history at an age of 5 Myr. In both cases the models are compared to this indices measured in this work, as shown in Figure 2, as well as the sub-sample of galaxies with measured indices from \citet{Leitherer11.1} that are purely star-forming galaxies (i.e., no active galactic nucleus contribution). Our data assumes KK04 metallicities for this comparison; the choice of strong line diagnostic has only a minimal effect on the overall agreement between the data and the models.

\begin{figure}[!ht]
	\includegraphics[width=\textwidth]{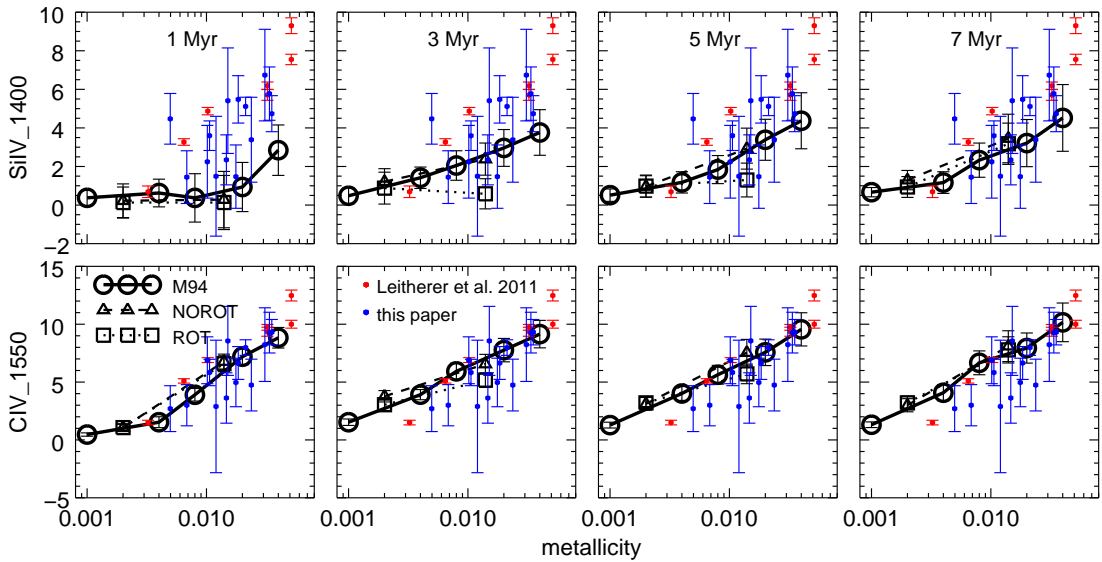}
	\caption{SiIV\_1400 (top) and CIV\_1550 (bottom) indices measured from the SB99 models (black), simulating an instantaneous burst star formation history at (from left to right) 1, 3, 5, and 7 Myr and adopting the M94 (circles), NOROT (triangles), and ROT (squares) stellar evolutionary tracks. The models are compared with the SiV\_1400 and CIV\_1550 indices measured from our HST data (blue) and the star-forming galaxies in \citet{Leitherer11.1} (red). All models assume a Kroupa IMF.}
	\label{fig:indices_SB99_inst}
\end{figure}

\begin{figure}[!ht]
		\includegraphics[width=\textwidth] {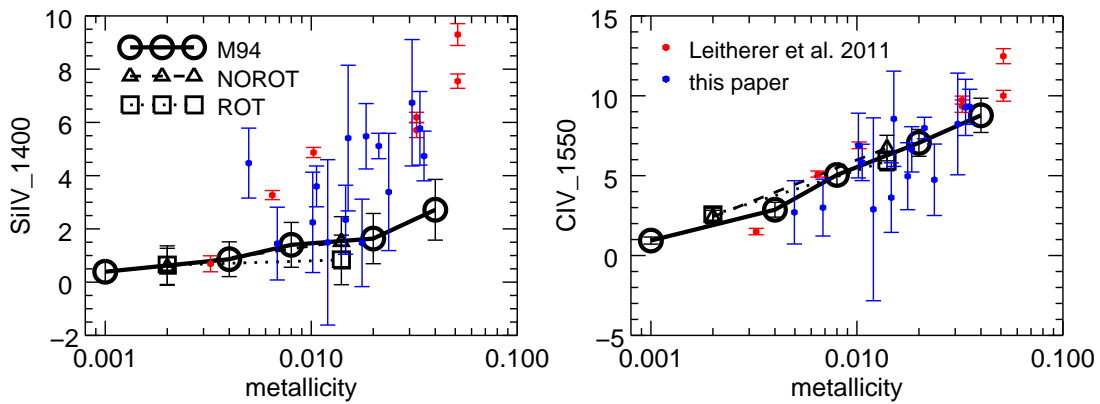}
		\caption{As in Figure 4, but for models simulating a continuous burst star formation history at an age of 5 Myr.}
	\label{fig:indices_SB99_cont_Leitherer}
\end{figure}

Figures 3 and 4 show generally good agreement between the observations and models. For the SiIV\_1400 index, the Starburst99 models generally underestimate the equivalent widths, particularly in the case of a continuous star formation history and the youngest ages of a single coeval stellar population (agreement improves at later ages). By contrast, in the case of the CIV\_1550 index there is strong agreement between the models and observations for both the individual index values and the slope of the relation as a whole across all ages and star formation histories. The underestimation of the SiIV\_1400 index by the models is expected; Starburst99 only simulates the stellar components of these indices, and as noted in \citet{Leitherer11.1} interstellar components could account for as much as $\sim$2/3 of the SiIV\_1400 equivalent width. Therefore, the equivalent widths predicted by Starburst99 for the SiIV\_1400 indices should be considered only as lower limits.

It is also clear that the ROT models predict weaker equivalent widths for both indices, which is explained by an overall shift to a hotter population of stars \citep[see][]{Levesque12.1,Leitherer14.1}. Si \textsc{iv} (and to a lesser extent C \textsc{iv}) are not the dominant ionization states of Si and C, and a hotter population of ionizing stars will lead to weaker lines as a result (in contrast, features of dominant species such as N \textsc{v} and O \textsc{vi} are stronger in models that include stellar rotation; \citealt{Leitherer14.1}).

\section{The $\beta_{18}$ Extinction Diagnostic}

In optical spectra, the H$\alpha$/H$\beta$ Balmer decrement is the most common diagnostic for quantifying extinction effects. By contrast, in the UV, the slope of the continuum is the most widely-used extinction diagnostic. The UV spectral slope, $\beta$, is determined by fitting the continuum to the function $f(\lambda) \propto \lambda^\beta$, where $f(\lambda)$ is the flux density, to a specified wavelength range. \citet{Calzetti94.1} consider both long-wavelength (1250--2800\AA) and short-wavelength (1250--1700\AA) definitions of $\beta$, finding robust correlations with $\tau^l_B$ (where $\tau^l_B$ = ln$\left(\frac{H\alpha/H\beta}{2.86}\right)$, taking 2.86 as the theoretical Balmer decrement from \citealt{Osterbrock89.1}). Similarly, \citet{Calzetti01.1} defines $\beta_{26}$, a fit spanning 1250--2600\AA, as a standardized measure of the UV slope and extinction diagnostic.

Here we now consider $\beta_{18}$, defined by \citet{Calzetti01.1} as the continuum slope from 1250--1800\AA, as a potential additional diagnostic for shorter wavelengths (and, consequently, higher redshifts). Figure 5 shows the relationship between $\beta_{18}$ and $\tau^l_B$ for our observed galaxies. In order to perform $\beta_{18}$ fits to the UV continuua we first mask all of the Leitherer indices which fall within the diagnostic wavelength range. In the HST COS spectra, the geocoronal O \textsc{i} 1302 emission lines were also masked (the geocoronal Ly$\alpha$ does not need masking at the redshifts of our galaxies; see Figure 1). We calculated H$\alpha$/H$\beta$ ratios for each of our galaxies using the existing optical spectra from \citet{Kong02.1}, and define $\tau^l_B$ identically to \citet{Calzetti94.1}. To remove any potential contamination from Galactic dust, we corrected both the raw line fluxes from \citet{Kong02.1} and our own UV spectra for Galactic reddening \citep{Schlegel98.1, Schlafly11.1}. Our values for $\beta_{18}$ and $\tau^l_B$ are given in Table 2, and our data show generally good agreement with the linear relationship determined in \citet{Calzetti94.1} for their definitions of a short-wavelength $\beta$ and $\tau^l_B$.

\begin{figure}[!ht]
	\includegraphics[width=\textwidth]{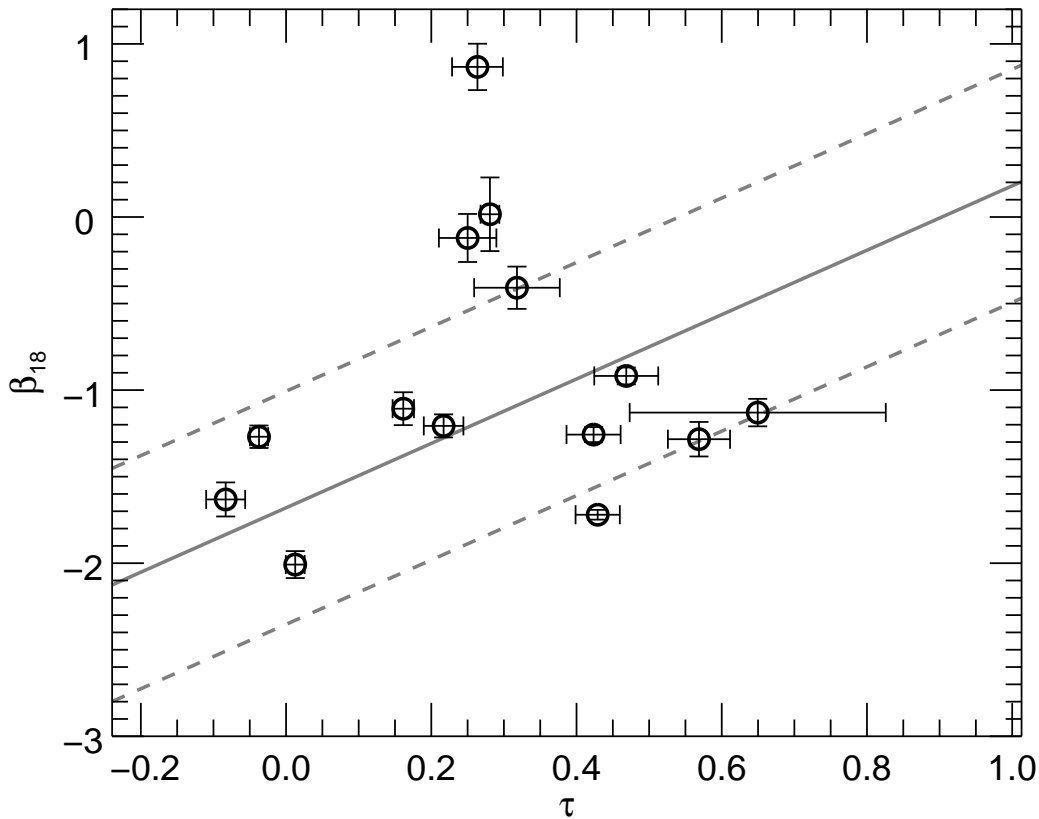}
	\caption{$\tau^l_B$ vs. $\beta_{18}$ for our sample of observed galaxies. Values from $\tau^l_B$ are determined from the optical data of \citet{Kong02.1}; values for $\beta_{18}$ are from our own HST spectra. The solid line is the best linear fit from \citet{Calzetti94.1}. The dashed lines represent the $\sim5\sigma$ upper and lower envelopes of their data.}
	\label{fig:BalmerBeta}
\end{figure}

We note that our definition of $\beta_{18}$ spans a slightly different wavelength range than the \citet{Calzetti94.1} ``short-wavelength" $\beta$ , covering 1250--1800\AA\ rather than the 1250--1700\AA. While the difference is slight this could account for some of the minor disagreements in Figure 5 between our data and the Calzetti relation (it is also worth noting that 16\% of the original \citealt{Calzetti94.1} data points also fall outside this relation). One galaxy, Haro 25, does fall significantly above the rest of the sample; however, there are discrepancies in the literature regarding the measured value of $\tau$ from the optical spectrum. Data from \citet{Kehrig04.1} and the H$\gamma$/H$\beta$ decrement from Kong et al.\ (2002) both imply a higher reddening than the $\tau$ determined here, which could partially improve this outlier's agreement. 

$\beta_{18}$ holds particular value as an extinction diagnostic at high redshift. Spectral coverage in the UV is often limited to a narrower wavelength range than that spanned by the $\beta_{26}$ diagnostic. At $z \gtrsim 3$ this regime is redshifted into the near-IR; as a result, a UV extinction diagnostic over a shorter wavelength range would be advantageous for studying extinction in high-redshift galaxies.

\subsection{Theoretical Values of $\beta_{18}$}

In Appendix A we provide tables of theoretical (dust-free) values for $\beta_{18}$ calculated from our Starburst99 models, to be used as references for determining extinction from both nearby and high-redshift galaxies using this definition of the UV continuum slope. Fully line-blanketed spectra were used to calculate the $\beta_{18}$ values, as opposed to using the true continuum. Our values are calculated for both instantaneous burst and continuous star-forming histories (at ages ranging from 0--20Myr in the instantaneous case and an equilibrium age of 5 Myr in the continuous case), all three stellar evolutionary tracks (M94, NOROT, and ROT), and all three IMFs simulated in our model grids (see Section 3).

The Starburst99 code is primarily based on stellar evolutionary synthesis, i.e., it is not a photoionization code. Therefore nebular physics is not considered and emission lines originating in HII regions and the general interstellar medium are not included in the SED. However, in the presence of hot ionizing stars {\it continuous nebular emission} can make a non-negligible contribution to the stellar emission and must be accounted for. Fortunately, nebular continuous emission is only weakly dependent on gas properties such as electron temperature ($T_e$) and electron density ($n_e$), and detailed photoionization modeling is not required for estimating its contribution. Therefore this process is included in Starburst99. Starburst99 in its default mode assumes all ionizing photons shortward of the Lyman break are absorbed in standard case B mode. We account for three continuous emission processes: bound-free + free-free emission of hydrogen; bound-free + free-free emission of neutral helium; and hydrogen two-photon emission. The absorption coefficients for these processes are taken from Tables 4--9 of \citet{Aller84.1}; they are supplemented by values at longer wavelengths from \citet{Ferland80.1}. Fixed values of $T_e = 10,000$ K and $n_e = 10^4$ cm$^{-3}$ are assumed (this assumption is non-critical for the parameter range of interest to this study).

A general discussion of the importance of the nebular relative to the stellar continuum can be found in \citet{Leitherer95.1}. Since the focus of this work is on the UV, we are providing more detail on the properties of the two-photon continuum, which was not explicitly covered by \citet{Leitherer95.1}. As a reminder, about 2/3 of all recombinations in the H atom lead to the $p$ levels, which then ultimately result in the H emission-line and bound-free spectrum. 1/3 of the recombinations go to the $2s$ level, which is meta-stable. Unless densities are extremely high ($>10^9$ cm$^{-3}$; \citealt{Drake81.1}) and collisional transitions to the $2p$ level occur, a transition from the $2s$ level to a virtual level takes place whose energy lies between those of the $1s$ and $2s$ levels. This process is called two-photon emission since it produces two photons with a combined energy equal to that of the $2s$ to $1s$ transition (i.e., Ly-$\alpha$). The probability distribution of the process is symmetrical about half the energy level difference of $2s$ to $1s$ transition, corresponding to a wavelength of 2431\AA. The peak emission of the two-photon spectrum is found to be at 1621\AA\ after accounting for the weighting by the photon energy.

Of the three nebular continuous emission processes included in Starburst99, the two-photon continuum by far dominates in the UV. The two-photon continuous absorption coefficients are an order of magnitude larger than the bound-free + free-free coefficients between 1200--2000\AA. \citet{Drake81.1} provide examples of the relative strengths of these processes. 

In order to assess the importance of the nebular contribution to the UV continuum, we computed two sets of models. One set is the Starburst99 default discussed above. This corresponds to case B in the radiation-bounded regime. Alternatively, we considered a second set having no nebular contribution. This could be a radiation-bounded nebula or a galaxy whose interstellar medium is porous due to the effects of stellar winds and supernovae which clear out cavities leading to the escape of ionizing photons \citep{Borthakur14.1}. Comparison of the two model sets with and without nebular continuum suggests that omission of the nebular continuum results in a bluer spectrum (more negative $\beta_{18}$). The reason for this behavior is the redder spectral slope of the nebular compared with the stellar continuum. Removal of the nebular component then leads to a bluer spectrum.  

\citet{Bouwens10.1} drew attention to the significance of the nebular continuum for the observed rest-frame UV spectra of galaxies at high redshift. The observed UV slopes are sometimes bluer than predicted by the most extreme theoretical models. Depending on dust reddening, metallicity, age, or initial mass function, the bluest slopes in the models have $\beta > -2.7$, whereas the observations in some galaxies suggest $\beta < -3$. Although observational uncertainties are large and part of the discrepancies may be ascribed to larger than assumed photometric errors \citep{Dunlop12.1}, there is tentative evidence for bluer observed slopes when larger samples are considered (e.g., \citealt{Jiang13.1}). Star-forming galaxies with redshifts approaching the epoch of cosmic reionization have been postulated to be leaking ionizing photons \citep{Kuhlen12.1}. If so, one would expect a transition from the radiation-bounded to the density-bounded regime, and therefore a diminished contribution by the nebular continuum to the UV spectrum.

Our two sets of $\beta_{18}$ values given in Appendix A are thus intended to bracket the two extreme cases of 0 and 100\% escape fraction ($f_{\rm esc}$) of ionizing photons. They are complementary to the models of \citet{Inoue11.1} who provided a grid of nebular models, including continuum models, using an older version of Starburst99 when evolution models with rotation had not yet been available.

The tables in Appendix A indicate several trends of the behavior of $\beta_{18}$. There is an overall evolution towards redder slopes with age, which is simply the result of on average lower stellar temperatures with older age. As an example, see Figure 6. The non-monotonic behavior around 4 to 6 Myr is due to the first appearance of luminous post-main-sequence stars whose somewhat cooler temperature leads to a redder UV continuum. These stars do not subsequently evolve into red supergiants but rather turn into hot Wolf-Rayet stars. Overall, the variation of $\beta_{18}$ with age is rather moderate, which of course is the reason for the suitability of the UV slope as a dust indicator.
 
 \begin{figure}[!ht]
	\includegraphics[width=0.5\textwidth]{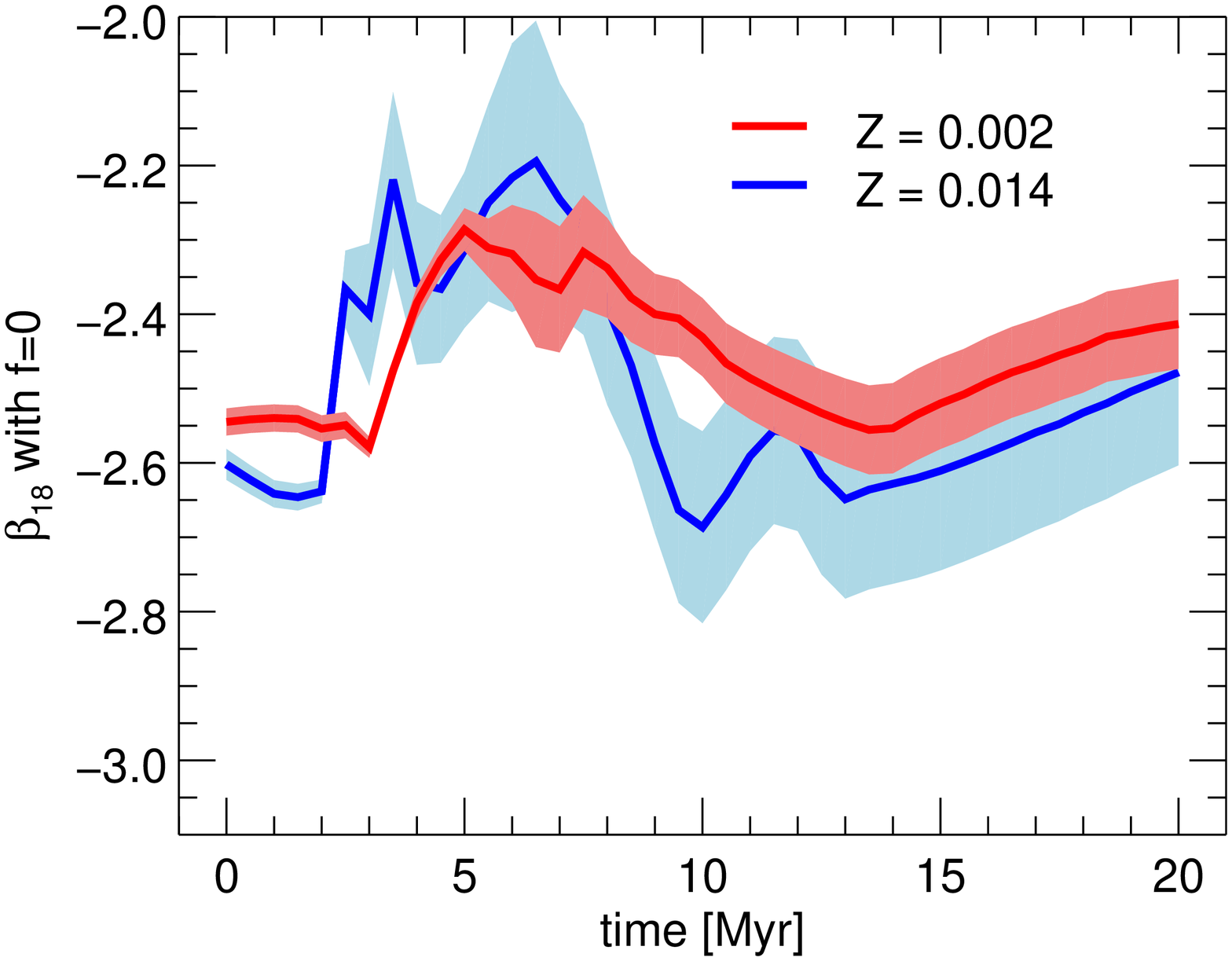}
	\includegraphics[width=0.5\textwidth]{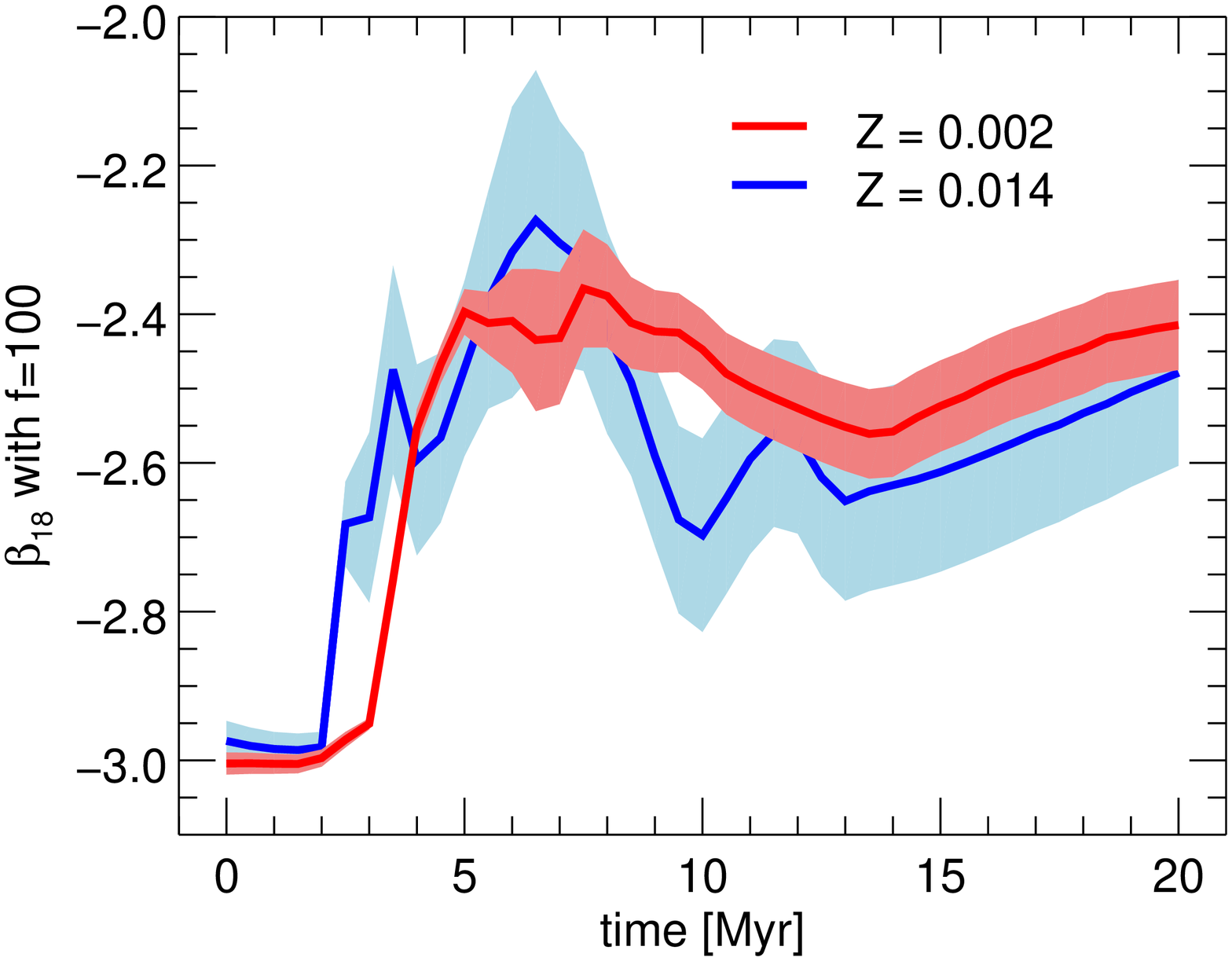}
	\caption{$\beta_{18}$ measured from the SB99 models, simulating an instantaneous burst star formation history with ROT evolutionary tracks and a Kroupa IMF. Metallicities are $Z=0.002$ (red) and $Z=0.014$ (blue). The two extreme cases of $f_{\rm esc} = 0$ (left) and 100 (right) are illustrated.}
	\label{fig:BetaTime}
\end{figure}

The three IMF choices considered here result in only minor differences in $\beta_{18}$. The Kroupa IMF has parameters ($\alpha_{low}=1.3$, $\alpha_{high}=2.3$). An IMF enriched in massive stars relative to the Kroupa IMF (1.3, 1.3) has a UV slope which is bluer by $\sim0.2$ than an IMF deficient of massive stars (1.3, 3.3). More extreme variations of $\beta_{18}$ are expected in the presence/absence of very massive stars with masses higher than, e.g., 150 M$_{\odot}$. These stars are not covered by the evolutionary tracks included in Starburst99.
 
The two extreme cases of $f_{\rm esc} = 0$ and 100 (all ionizing photons absorbed or escaping, respectively) have a rather significant effect on $\beta_{18}$ for ages $<10$ Myr. For ages and IMF values favoring the production of ionizing radiation, the UV slopes with $f_{\rm esc} = 100$ are bluer by $\sim0.5$. This behavior is easily understandable from the previous discussion of the nebular continuum: the nebular continuum is dominated by the two-photon process whose slope is redder than that of the stellar continuum.
 
Metallicity has a relatively minor impact on $\beta_{18}$. While individual metal-poor stars have bluer colors than metal-rich stars, stellar evolution at high metallicity favors the presence of hot Wolf-Rayet stars, which compensates for the color effect in individual stars. As a result, the $\beta_{18}$ values at various metallicities are not very different. We remind the reader that this argument only applies to models for single-star evolution. If significant numbers of Wolf-Rayet stars at low metallicity are produced in binaries, $\beta_{18}$ may become bluer as well.
 
Models with rotation generally result in more negative $\beta_{18}$ values than non-rotating models. The effect is less pronounced for $f_{\rm esc} = 0$ because rotation leads to hotter stars which are bluer on the one hand but are also responsible for a stronger nebular continuum on the other.

\section{Discussion and future work}

We have demonstrated that there is a statistically robust positive linear correlation with metallicity for the UV stellar wind indices SiIV\_1400 and CIV\_1550. This correlation is supported both by multi-wavelength spectroscopic observations of star-forming galaxies and by the predictions of the Starburst99 stellar population synthesis models. This suggests that parameterizations of UV metallicity diagnostics based on synthetic galaxy spectra will offer a well-calibrated suite of new global metallicity diagnostics that can be applied to spectra of star-forming galaxies at high redshifts.

In addition, the clear correlation between the UV and optical metallicity diagnostics has important implications for future studies of star-forming galaxy metallicities across a broad range of redshifts. While a larger and higher-S/N multi-wavelength observational sample is required to fully sample the parameter space of ISM properties and diagnostics in star-forming galaxies (including star formation rate, young stellar population age, stellar mass, and extinction), our results demonstrate that it will be possible to compare metallicities determined from both UV and optical spectra through conversions similar to those calculated by \citet{Kewley08.1} for optical diagnostics alone. This will offer the first quantitative means of unifying metallicity measurements for star-forming galaxies across multiple wavelengths, and consequently across a much broader range of redshifts.

Similarly, combining the ionizing SED outputs of the full grid of Starburst99 models (including those accommodating the effects of stellar rotation, e.g. \citealt{Levesque12.1,Leitherer14.1}) with a photoionization code such as MAPPINGS \citep[e.g.][]{Sutherland93.1,Groves04.1} or CLOUDY \citep{Ferland98.1} would allow us to determine theoretical parameterizations for indices with strong or dominant contributions from interstellar lines and directly compare these to predicted fluxes for optical emission lines. This includes the interstellar components of the SiIV\_1400 and CIV\_1550 indices as well as other far-UV diagnostic indices that are primarily interstellar such as CII\_1330 and AlII\_1670.

We have also determined theoretical values of $\beta_{18}$ that can be used for determining extinction in rest-frame UV spectra of star-forming galaxies, assuming the two extremes of 0\% and 100\% $f_{\rm esc}$ to facilitate their use with observations of star-forming galaxies in both the nearby and high-redshift universe. Following previous work by \citet{Calzetti94.1} and \citet{Calzetti01.1}, $\beta_{18}$ shows a similar correlation with optical extinction diagnostics and is particularly valuable for high-redshift observations. In the future, a combination of stellar population synthesis and photoionization models would also allow for a valuable theoretical comparison between the extinction diagnostics $\beta_{18}$ {\it and} $\tau^l_B$, allowing us to directly probe the effects of other galaxy properties (such as stellar population age, ionization parameter, and electron density) on these diagnostics and improving their utility at high redshift.

\acknowledgements This paper was based on observations made with the NASA/ESA Hubble Space Telescope associated with program 13481; support for this program was provided by NASA through a grant from the Space Telescope Science Institute. EML is additionally supported by NASA through Hubble Fellowship grant  number HST-HF-51324.01-A from the Space Telescope Science Institute. The Space Telescope Science Institute is operated by the Association of Universities for Research in Astronomy, Incorporated, under NASA contract NAS5-26555.

\bibliographystyle{apj}

\begin{thebibliography}{}
\expandafter\ifx\csname natexlab\endcsname\relax\def\natexlab#1{#1}\fi

\bibitem[{{Aller}(1984)}]{Aller84.1}
{Aller}, L.~H., ed. 1984, Astrophysics and Space Science Library, Vol. 112,
  {Physics of thermal gaseous nebulae}

\bibitem[{{Borthakur} {et~al.}(2014){Borthakur}, {Heckman}, {Leitherer}, \&
  {Overzier}}]{Borthakur14.1}
{Borthakur}, S., {Heckman}, T.~M., {Leitherer}, C., \& {Overzier}, R.~A. 2014,
  Science, 346, 216

\bibitem[{{Bouwens} {et~al.}(2010){Bouwens}, {Illingworth}, {Oesch}, {Trenti},
  {Stiavelli}, {Carollo}, {Franx}, {van Dokkum}, {Labb{\'e}}, \&
  {Magee}}]{Bouwens10.1}
{Bouwens}, R.~J., {Illingworth}, G.~D., {Oesch}, P.~A., {et~al.} 2010, \apjl,
  708, L69

\bibitem[{{Burstein} {et~al.}(1984){Burstein}, {Faber}, {Gaskell}, \&
  {Krumm}}]{Burstein84.1}
{Burstein}, D., {Faber}, S.~M., {Gaskell}, C.~M., \& {Krumm}, N. 1984, \apj,
  287, 586

\bibitem[{{Calzetti}(2001)}]{Calzetti01.1}
{Calzetti}, D. 2001, \pasp, 113, 1449

\bibitem[{{Calzetti} \& {Heckman}(1999)}]{Calzetti99.1}
{Calzetti}, D., \& {Heckman}, T.~M. 1999, \apj, 519, 27

\bibitem[{{Calzetti} {et~al.}(1994){Calzetti}, {Kinney}, \&
  {Storchi-Bergmann}}]{Calzetti94.1}
{Calzetti}, D., {Kinney}, A.~L., \& {Storchi-Bergmann}, T. 1994, \apj, 429, 582

\bibitem[{{Danforth} {et~al.}(2010){Danforth}, {Keeney}, {Stocke}, {Shull}, \&
  {Yao}}]{Danforth10.1}
{Danforth}, C.~W., {Keeney}, B.~A., {Stocke}, J.~T., {Shull}, J.~M., \& {Yao},
  Y. 2010, \apj, 720, 976

\bibitem[{{Denicol{\'o}} {et~al.}(2002){Denicol{\'o}}, {Terlevich}, \&
  {Terlevich}}]{Denicolo02.1}
{Denicol{\'o}}, G., {Terlevich}, R., \& {Terlevich}, E. 2002, \mnras, 330, 69

\bibitem[{{Drake} \& {Ulrich}(1981)}]{Drake81.1}
{Drake}, S.~A., \& {Ulrich}, R.~K. 1981, \apj, 248, 380

\bibitem[{{Dunlop} {et~al.}(2012){Dunlop}, {McLure}, {Robertson}, {Ellis},
  {Stark}, {Cirasuolo}, \& {de Ravel}}]{Dunlop12.1}
{Dunlop}, J.~S., {McLure}, R.~J., {Robertson}, B.~E., {et~al.} 2012, \mnras,
  420, 901

\bibitem[{{Ekstr{\"o}m} {et~al.}(2012){Ekstr{\"o}m}, {Georgy}, {Eggenberger},
  {Meynet}, {Mowlavi}, {Wyttenbach}, {Granada}, {Decressin}, {Hirschi},
  {Frischknecht}, {Charbonnel}, \& {Maeder}}]{Ekstrom12.1}
{Ekstr{\"o}m}, S., {Georgy}, C., {Eggenberger}, P., {et~al.} 2012, \aap, 537,
  A146

\bibitem[{{Faber} {et~al.}(1985){Faber}, {Friel}, {Burstein}, \&
  {Gaskell}}]{Faber85.1}
{Faber}, S.~M., {Friel}, E.~D., {Burstein}, D., \& {Gaskell}, C.~M. 1985,
  \apjs, 57, 711

\bibitem[{{Ferland}(1980)}]{Ferland80.1}
{Ferland}, G.~J. 1980, \pasp, 92, 596

\bibitem[{{Ferland} {et~al.}(1998){Ferland}, {Korista}, {Verner}, {Ferguson},
  {Kingdon}, \& {Verner}}]{Ferland98.1}
{Ferland}, G.~J., {Korista}, K.~T., {Verner}, D.~A., {et~al.} 1998, \pasp, 110,
  761

\bibitem[{{Georgy} {et~al.}(2013){Georgy}, {Ekstr{\"o}m}, {Eggenberger},
  {Meynet}, {Haemmerl{\'e}}, {Maeder}, {Granada}, {Groh}, {Hirschi}, {Mowlavi},
  {Yusof}, {Charbonnel}, {Decressin}, \& {Barblan}}]{Georgy13.1}
{Georgy}, C., {Ekstr{\"o}m}, S., {Eggenberger}, P., {et~al.} 2013, \aap, 558,
  A103

\bibitem[{{Gonz{\'a}lez Delgado} {et~al.}(1998){Gonz{\'a}lez Delgado},
  {Leitherer}, {Heckman}, {Lowenthal}, {Ferguson}, \&
  {Robert}}]{GonzalezDelgado98.1}
{Gonz{\'a}lez Delgado}, R.~M., {Leitherer}, C., {Heckman}, T., {et~al.} 1998,
  \apj, 495, 698

\bibitem[{{Groves} {et~al.}(2004){Groves}, {Dopita}, \&
  {Sutherland}}]{Groves04.1}
{Groves}, B.~A., {Dopita}, M.~A., \& {Sutherland}, R.~S. 2004, \apjs, 153, 9

\bibitem[{{Hamann} \& {Gr{\"a}fener}(2004)}]{Hamann04.1}
{Hamann}, W.-R., \& {Gr{\"a}fener}, G. 2004, \aap, 427, 697

\bibitem[{{Heckman} {et~al.}(1998){Heckman}, {Robert}, {Leitherer}, {Garnett},
  \& {van der Rydt}}]{Heckman98.1}
{Heckman}, T.~M., {Robert}, C., {Leitherer}, C., {Garnett}, D.~R., \& {van der
  Rydt}, F. 1998, \apj, 503, 646

\bibitem[{{Hopkins} \& {Beacom}(2006)}]{Hopkins06.1}
{Hopkins}, A.~M., \& {Beacom}, J.~F. 2006, \apj, 651, 142

\bibitem[{{Huang} {et~al.}(2010){Huang}, {Gies}, \& {McSwain}}]{Huang10.1}
{Huang}, W., {Gies}, D.~R., \& {McSwain}, M.~V. 2010, \apj, 722, 605

\bibitem[{{Inoue}(2011)}]{Inoue11.1}
{Inoue}, A.~K. 2011, \mnras, 415, 2920

\bibitem[{{Jiang} {et~al.}(2013){Jiang}, {Egami}, {Mechtley}, {Fan}, {Cohen},
  {Windhorst}, {Dav{\'e}}, {Finlator}, {Kashikawa}, {Ouchi}, \&
  {Shimasaku}}]{Jiang13.1}
{Jiang}, L., {Egami}, E., {Mechtley}, M., {et~al.} 2013, \apj, 772, 99

\bibitem[{{Kehrig} {et~al.}(2004){Kehrig}, {Telles}, \&
  {Cuisinier}}]{Kehrig04.1}
{Kehrig}, C., {Telles}, E., \& {Cuisinier}, F. 2004, \aj, 128, 1141

\bibitem[{{Kewley} \& {Dopita}(2002)}]{Kewley02.1}
{Kewley}, L.~J., \& {Dopita}, M.~A. 2002, \apjs, 142, 35

\bibitem[{{Kewley} {et~al.}(2001){Kewley}, {Dopita}, {Sutherland}, {Heisler},
  \& {Trevena}}]{Kewley01.1}
{Kewley}, L.~J., {Dopita}, M.~A., {Sutherland}, R.~S., {Heisler}, C.~A., \&
  {Trevena}, J. 2001, \apj, 556, 121

\bibitem[{{Kewley} \& {Ellison}(2008)}]{Kewley08.1}
{Kewley}, L.~J., \& {Ellison}, S.~L. 2008, \apj, 681, 1183

\bibitem[{{Kewley} {et~al.}(2006){Kewley}, {Groves}, {Kauffmann}, \&
  {Heckman}}]{Kewley06.1}
{Kewley}, L.~J., {Groves}, B., {Kauffmann}, G., \& {Heckman}, T. 2006, \mnras,
  372, 961

\bibitem[{{Kobulnicky} \& {Kewley}(2004)}]{Kobulnicky04.1}
{Kobulnicky}, H.~A., \& {Kewley}, L.~J. 2004, \apj, 617, 240

\bibitem[{{Kong} \& {Cheng}(2002)}]{Kong02.2}
{Kong}, X., \& {Cheng}, F.~Z. 2002, \aap, 389, 845

\bibitem[{{Kong} {et~al.}(2002){Kong}, {Cheng}, {Weiss}, \&
  {Charlot}}]{Kong02.1}
{Kong}, X., {Cheng}, F.~Z., {Weiss}, A., \& {Charlot}, S. 2002, \aap, 396, 503

\bibitem[{{Kroupa}(2001)}]{Kroupa01.1}
{Kroupa}, P. 2001, \mnras, 322, 231

\bibitem[{{Kuhlen} \& {Faucher-Gigu{\`e}re}(2012)}]{Kuhlen12.1}
{Kuhlen}, M., \& {Faucher-Gigu{\`e}re}, C.-A. 2012, \mnras, 423, 862

\bibitem[{{Leitherer} \& {Chen}(2009)}]{Leitherer09.1}
{Leitherer}, C., \& {Chen}, J. 2009, \na, 14, 356

\bibitem[{{Leitherer} {et~al.}(2014){Leitherer}, {Ekstr{\"o}m}, {Meynet},
  {Schaerer}, {Agienko}, \& {Levesque}}]{Leitherer14.1}
{Leitherer}, C., {Ekstr{\"o}m}, S., {Meynet}, G., {et~al.} 2014, \apjs, 212, 14

\bibitem[{{Leitherer} \& {Heckman}(1995)}]{Leitherer95.1}
{Leitherer}, C., \& {Heckman}, T.~M. 1995, \apjs, 96, 9

\bibitem[{{Leitherer} {et~al.}(2001){Leitherer}, {Le{\~a}o}, {Heckman},
  {Lennon}, {Pettini}, \& {Robert}}]{Leitherer01.1}
{Leitherer}, C., {Le{\~a}o}, J.~R.~S., {Heckman}, T.~M., {et~al.} 2001, \apj,
  550, 724

\bibitem[{{Leitherer} {et~al.}(2010){Leitherer}, {Ortiz Ot{\'a}lvaro},
  {Bresolin}, {Kudritzki}, {Lo Faro}, {Pauldrach}, {Pettini}, \&
  {Rix}}]{Leitherer10.1}
{Leitherer}, C., {Ortiz Ot{\'a}lvaro}, P.~A., {Bresolin}, F., {et~al.} 2010,
  \apjs, 189, 309

\bibitem[{{Leitherer} {et~al.}(2011){Leitherer}, {Tremonti}, {Heckman}, \&
  {Calzetti}}]{Leitherer11.1}
{Leitherer}, C., {Tremonti}, C.~A., {Heckman}, T.~M., \& {Calzetti}, D. 2011,
  \aj, 141, 37

\bibitem[{{Leitherer} {et~al.}(1999){Leitherer}, {Schaerer}, {Goldader},
  {Delgado}, {Robert}, {Kune}, {de Mello}, {Devost}, \&
  {Heckman}}]{Leitherer99.1}
{Leitherer}, C., {Schaerer}, D., {Goldader}, J.~D., {et~al.} 1999, \apjs, 123,
  3

\bibitem[{{Levesque} {et~al.}(2010){Levesque}, {Berger}, {Kewley}, \&
  {Bagley}}]{Levesque10.1}
{Levesque}, E.~M., {Berger}, E., {Kewley}, L.~J., \& {Bagley}, M.~M. 2010, \aj,
  139, 694

\bibitem[{{Levesque} {et~al.}(2012){Levesque}, {Leitherer}, {Ekstrom},
  {Meynet}, \& {Schaerer}}]{Levesque12.1}
{Levesque}, E.~M., {Leitherer}, C., {Ekstrom}, S., {Meynet}, G., \& {Schaerer},
  D. 2012, \apj, 751, 67

\bibitem[{{Meynet} {et~al.}(1994){Meynet}, {Maeder}, {Schaller}, {Schaerer}, \&
  {Charbonnel}}]{Meynet94.1}
{Meynet}, G., {Maeder}, A., {Schaller}, G., {Schaerer}, D., \& {Charbonnel}, C.
  1994, \aaps, 103, 97

\bibitem[{{Osterbrock}(1989)}]{Osterbrock89.1}
{Osterbrock}, D.~E. 1989, {Astrophysics of gaseous nebulae and active galactic
  nuclei} (University Science Books)

\bibitem[{{Papovich} {et~al.}(2001){Papovich}, {Dickinson}, \&
  {Ferguson}}]{Papovich01.1}
{Papovich}, C., {Dickinson}, M., \& {Ferguson}, H.~C. 2001, \apj, 559, 620

\bibitem[{{Pauldrach} {et~al.}(2001){Pauldrach}, {Hoffmann}, \&
  {Lennon}}]{Pauldrach01.1}
{Pauldrach}, A.~W.~A., {Hoffmann}, T.~L., \& {Lennon}, M. 2001, \aap, 375, 161

\bibitem[{{Pettini} \& {Pagel}(2004)}]{Pettini04.1}
{Pettini}, M., \& {Pagel}, B.~E.~J. 2004, \mnras, 348, L59

\bibitem[{{Rix} {et~al.}(2004){Rix}, {Pettini}, {Leitherer}, {Bresolin},
  {Kudritzki}, \& {Steidel}}]{Rix04.1}
{Rix}, S.~A., {Pettini}, M., {Leitherer}, C., {et~al.} 2004, \apj, 615, 98

\bibitem[{{Schlafly} \& {Finkbeiner}(2011)}]{Schlafly11.1}
{Schlafly}, E.~F., \& {Finkbeiner}, D.~P. 2011, \apj, 737, 103

\bibitem[{{Schlegel} {et~al.}(1998){Schlegel}, {Finkbeiner}, \&
  {Davis}}]{Schlegel98.1}
{Schlegel}, D.~J., {Finkbeiner}, D.~P., \& {Davis}, M. 1998, \apj, 500, 525

\bibitem[{{Shapley} {et~al.}(2003){Shapley}, {Steidel}, {Pettini}, \&
  {Adelberger}}]{Shapley03.1}
{Shapley}, A.~E., {Steidel}, C.~C., {Pettini}, M., \& {Adelberger}, K.~L. 2003,
  \apj, 588, 65

\bibitem[{{Sutherland} \& {Dopita}(1993)}]{Sutherland93.1}
{Sutherland}, R.~S., \& {Dopita}, M.~A. 1993, \apjs, 88, 253

\bibitem[{{V{\'a}zquez} \& {Leitherer}(2005)}]{Vazquez05.1}
{V{\'a}zquez}, G.~A., \& {Leitherer}, C. 2005, \apj, 621, 695

\bibitem[{{Witt} {et~al.}(1992){Witt}, {Thronson}, \& {Capuano}}]{Witt92.1}
{Witt}, A.~N., {Thronson}, Jr., H.~A., \& {Capuano}, Jr., J.~M. 1992, \apj,
  393, 611

\end{thebibliography}

\appendix
\setcounter{table}{0}
\renewcommand{\thetable}{A\arabic{table}}

\section{$\beta_{18}$ in Star-Forming Galaxy Models}

\begin{landscape}
\begin{deluxetable}{l c c c c c | c c c c c | c c c c c}
\tabletypesize{\scriptsize}
\tablewidth{0pt}
\tablecaption{$\beta_{18}$ in Starburst99 Models with M94 Evolutionary Tracks, $f_{\rm esc}=0$} 
\tablecolumns{16}
\label{tab:high}
\tablehead{
\colhead{Age}
& \multicolumn{5}{c}{IMF = 1.3,1.3}
& \multicolumn{5}{c}{IMF = 1.3,2.3}
& \multicolumn{5}{c}{IMF = 1.3,3.3} \\ \cline{2-6} \cline{7-11} \cline{12-16}
\colhead{(Myr)}
&\colhead{0.001}
&\colhead{0.004}
&\colhead{0.008}
&\colhead{0.020}
&\colhead{0.040}
&\colhead{0.001}
&\colhead{0.004}
&\colhead{0.008}
&\colhead{0.020}
&\colhead{0.040}
&\colhead{0.001}
&\colhead{0.004}
&\colhead{0.008}
&\colhead{0.020}
&\colhead{0.040}
}
\startdata
       0.0   &  -2.3582 &  -2.4315 &  -2.4681 &  -2.5303 &  -2.6174   &     -2.4936 &  -2.5563 &  -2.5763 &  -2.5999 &  -2.6655    &    -2.6285 &  -2.6516 &  -2.6432  & -2.5977 &  -2.6261 \\
       2.0   &  -2.3579 &  -2.4989 &  -2.5528 &  -2.5164 &  -2.4756   &     -2.4870 &  -2.5813 &  -2.6138 &  -2.5888 &  -2.5010    &    -2.6258 &  -2.6460 &  -2.6365  & -2.6114 &  -2.5195 \\
       4.0   &  -1.9900 &  -2.3550 &  -2.1951 &  -2.0406 &  -1.7666   &     -2.2533 &  -2.4907 &  -2.3632 &  -2.1892 &  -1.9992    &    -2.5178 &  -2.5926 &  -2.5014  & -2.3348 &  -2.2026 \\
       6.0   &  -2.3226 &  -2.1695 &  -2.1999 &  -2.1696 &  -2.4203   &     -2.4730 &  -2.3546 &  -2.3646 &  -2.3147 &  -2.4869    &    -2.5839 &  -2.4851 &  -2.4670  & -2.3974 &  -2.4741 \\
       8.0   &  -2.4624 &  -2.4085 &  -2.7041 &  -2.6920 &  -2.6077   &     -2.5612 &  -2.5170 &  -2.7184 &  -2.6844 &  -2.5892    &    -2.6103 &  -2.5528 &  -2.6493  & -2.5932 &  -2.4937 \\
      10.0   &  -2.4620 &  -2.4402 &  -2.7933 &  -2.7155 &  -2.5876   &     -2.5528 &  -2.5290 &  -2.7694 &  -2.6807 &  -2.5460    &    -2.5864 &  -2.5392 &  -2.6561  & -2.5602 &  -2.4281 \\
      12.0   &  -2.4640 &  -2.4698 &  -2.7935 &  -2.6916 &  -2.5423   &     -2.5471 &  -2.5368 &  -2.7524 &  -2.6398 &  -2.4853    &    -2.5651 &  -2.5193 &  -2.6203  & -2.5031 &  -2.3540 \\
      14.0   &  -2.4446 &  -2.4952 &  -2.7643 &  -2.6375 &  -2.4646   &     -2.5227 &  -2.5396 &  -2.7124 &  -2.5776 &  -2.4009    &    -2.5319 &  -2.4963 &  -2.5703  & -2.4351 &  -2.2652 \\
      16.0   &  -2.4502 &  -2.4944 &  -2.6345 &  -2.3670 &  -2.3853   &     -2.5173 &  -2.5238 &  -2.6003 &  -2.3579 &  -2.3176    &    -2.5114 &  -2.4655 &  -2.4767  & -2.2663 &  -2.1803 \\
      18.0   &  -2.4352 &  -2.3870 &  -2.3452 &  -2.2390 &  -2.3084   &     -2.4940 &  -2.4275 &  -2.3702 &  -2.2421 &  -2.2375    &    -2.4795 &  -2.3845 &  -2.3117  & -2.1665 &  -2.0993 \\
      20.0   &  -2.4114 &  -2.3583 &  -2.3309 &  -2.2279 &  -2.2330   &     -2.4648 &  -2.3915 &  -2.3434 &  -2.2154 &  -2.1597    &    -2.4447 &  -2.3428 &  -2.2744  & -2.1254 &  -2.0207 \\ \hline
5.0 (cont)   &  -2.2595 &  -2.3961 &  -2.3491 &  -2.3757 &  -2.2533   &     -2.4101 &  -2.5119 &  -2.4617 &  -2.4431 &  -2.3015    &    -2.5851 &  -2.6115 &  -2.5573  & -2.4966 &  -2.3722 \\
 \bottomrule
\enddata
\end{deluxetable}

\begin{deluxetable}{l c c c c c | c c c c c | c c c c c}
\tabletypesize{\scriptsize}
\tablewidth{0pc}
\tablecaption{$\beta_{18}$ in Starburst99 Models with M94 Evolutionary Tracks, $f_{\rm esc}=100$} \tablecolumns{16}
\label{tab:high}
\tablehead{
\colhead{Age}
& \multicolumn{5}{c}{IMF = 1.3,1.3}
& \multicolumn{5}{c}{IMF = 1.3,2.3}
& \multicolumn{5}{c}{IMF = 1.3,3.3} \\ \cline{2-6} \cline{7-11} \cline{12-16}
\colhead{(Myr)}
&\colhead{0.001}
&\colhead{0.004}
&\colhead{0.008}
&\colhead{0.020}
&\colhead{0.040}
&\colhead{0.001}
&\colhead{0.004}
&\colhead{0.008}
&\colhead{0.020}
&\colhead{0.040}
&\colhead{0.001}
&\colhead{0.004}
&\colhead{0.008}
&\colhead{0.020}
&\colhead{0.040}
}
\startdata
       0.0   &  -3.0116 &  -3.0182 &  -3.0274  & -3.0340 &  -3.0690   &     -2.9980 &  -3.0020 &  -2.9979 &  -2.9767 &  -2.9989   &     -2.8785  & -2.8718 &  -2.8553 &  -2.7920  & -2.7996 \\
       2.0   &  -3.0053 &  -2.9722 &  -2.9671  & -2.7910 &  -2.6393   &     -2.9944 &  -2.9667 &  -2.9464 &  -2.8199 &  -2.6421   &     -2.8914  & -2.8603 &  -2.8247 &  -2.7521  & -2.6135 \\
       4.0   &  -2.1098 &  -2.4927 &  -2.3095  & -2.1159 &  -1.8247   &     -2.4001 &  -2.6240 &  -2.4727 &  -2.2560 &  -2.0443   &     -2.6481  & -2.6930 &  -2.5851 &  -2.3856  & -2.2338 \\
       6.0   &  -2.4094 &  -2.2194 &  -2.2319  & -2.1838 &  -2.4660   &     -2.5616 &  -2.4063 &  -2.3972 &  -2.3295 &  -2.5198   &     -2.6549  & -2.5274 &  -2.4937 &  -2.4103  & -2.4945 \\
       8.0   &  -2.5034 &  -2.4293 &  -2.7203  & -2.6993 &  -2.6113   &     -2.6016 &  -2.5370 &  -2.7328 &  -2.6910 &  -2.5925   &     -2.6429  & -2.5686 &  -2.6600 &  -2.5983  & -2.4964 \\
      10.0   &  -2.4773 &  -2.4479 &  -2.8002  & -2.7189 &  -2.5893   &     -2.5679 &  -2.5365 &  -2.7755 &  -2.6838 &  -2.5476   &     -2.5988  & -2.5453 &  -2.6606 &  -2.5626  & -2.4293 \\
      12.0   &  -2.4708 &  -2.4738 &  -2.7972  & -2.6934 &  -2.5432   &     -2.5538 &  -2.5406 &  -2.7556 &  -2.6415 &  -2.4861   &     -2.5707  & -2.5223 &  -2.6227 &  -2.5043  & -2.3546 \\
      14.0   &  -2.4483 &  -2.4978 &  -2.7666  & -2.6387 &  -2.4651   &     -2.5263 &  -2.5419 &  -2.7144 &  -2.5786 &  -2.4014   &     -2.5349  & -2.4981 &  -2.5718 &  -2.4359  & -2.2656 \\
      16.0   &  -2.4525 &  -2.4961 &  -2.6360  & -2.3677 &  -2.3857   &     -2.5196 &  -2.5253 &  -2.6016 &  -2.3584 &  -2.3179   &     -2.5132  & -2.4667 &  -2.4777 &  -2.2667  & -2.1805 \\
      18.0   &  -2.4368 &  -2.3880 &  -2.3459  & -2.2393 &  -2.3086   &     -2.4956 &  -2.4285 &  -2.3709 &  -2.2425 &  -2.2378   &     -2.4808  & -2.3853 &  -2.3123 &  -2.1668  & -2.0995 \\
      20.0   &  -2.4125 &  -2.3590 &  -2.3314  & -2.2282 &  -2.2332   &     -2.4659 &  -2.3921 &  -2.3439 &  -2.2156 &  -2.1598   &     -2.4457  & -2.3434 &  -2.2748 &  -2.1256  & -2.0208 \\ \hline
5.0 (cont)   &  -2.6150 &  -2.6999 &  -2.6211  & -2.6008 &  -2.4184   &     -2.7258 &  -2.7688 &  -2.6842 &  -2.6152 &  -2.4231   &     -2.7844  & -2.7695 &  -2.6938 &  -2.5996  & -2.4448 \\ \bottomrule
\enddata
\end{deluxetable}

\end{landscape}

\begin{deluxetable}{l c c | c c | c c}
\tabletypesize{\scriptsize}
\tablewidth{0pc}
\tablecaption{$\beta_{18}$ in Starburst99 Models with NOROT Evolutionary Tracks, $f_{\rm esc}=0$} 
\tablecolumns{7}
\label{tab:v00}
\tablehead{
\colhead{Age}
& \multicolumn{2}{c}{IMF = 1.3,1.3}
& \multicolumn{2}{c}{IMF = 1.3,2.3}
& \multicolumn{2}{c}{IMF = 1.3,3.3} \\ \cline{2-3} \cline{4-5} \cline{6-7}
\colhead{(Myr)}
&\colhead{0.002}
&\colhead{0.014}
&\colhead{0.002}
&\colhead{0.014}
&\colhead{0.002}
&\colhead{0.014}
}
\startdata
       0.0   &  -2.4064 &  -2.5104    &    -2.5245  & -2.5905     &   -2.6395 &  -2.6198 \\
       2.0   &  -2.4144 &  -2.5906    &    -2.5271  & -2.6468     &   -2.6431 &  -2.6423 \\
       4.0   &  -2.0675 &  -2.0281    &    -2.2988  & -2.1726     &   -2.5236 &  -2.3276 \\
       6.0   &  -2.2837 &  -2.1980    &    -2.4284  & -2.3378     &   -2.5363 &  -2.4155 \\
       8.0   &  -2.4082 &  -2.6100    &    -2.5132  & -2.6303     &   -2.5630 &  -2.5599 \\
      10.0   &  -2.4129 &  -2.6489    &    -2.5098  & -2.6357     &   -2.5378 &  -2.5282 \\
      12.0   &  -2.4061 &  -2.6926    &    -2.4928  & -2.6423     &   -2.5039 &  -2.4993 \\
      14.0   &  -2.3747 &  -2.6771    &    -2.4548  & -2.6068     &   -2.4569 &  -2.4466 \\
      16.0   &  -2.3310 &  -2.6169    &    -2.4080  & -2.5405     &   -2.4058 &  -2.3770 \\
      18.0   &  -2.3233 &  -2.5536    &    -2.3907  & -2.4739     &   -2.3765 &  -2.3097 \\
      20.0   &  -2.3581 &  -2.4955    &    -2.4071  & -2.4131     &   -2.3723 &  -2.2489 \\ \hline
5.0 (cont)   &  -2.3203 &  -2.3289    &    -2.4545  & -2.4249     &   -2.6024 &  -2.5026 \\ \bottomrule
\enddata
\end{deluxetable}

\begin{deluxetable}{l c c | c c | c c}
\tabletypesize{\scriptsize}
\tablewidth{0pc}
\tablecaption{$\beta_{18}$ in Starburst99 Models with NOROT Evolutionary Tracks, $f_{\rm esc}=100$} 
\tablecolumns{7}
\label{tab:v00}
\tablehead{
\colhead{Age}
& \multicolumn{2}{c}{IMF = 1.3,1.3}
& \multicolumn{2}{c}{IMF = 1.3,2.3}
& \multicolumn{2}{c}{IMF = 1.3,3.3} \\ \cline{2-3} \cline{4-5} \cline{6-7}
\colhead{(Myr)}
&\colhead{0.002}
&\colhead{0.014}
&\colhead{0.002}
&\colhead{0.014}
&\colhead{0.002}
&\colhead{0.014}
}
\startdata
       0.0  &   -3.0202 &  -3.0367    &    -3.0071 &  -2.9926     &   -2.8862  & -2.8306 \\
       2.0  &   -3.0074 &  -2.9072    &    -3.0017 &  -2.9155     &   -2.9001  & -2.8042 \\
       4.0  &   -2.1767 &  -2.1083    &    -2.4284 &  -2.2469     &   -2.6385  & -2.3862 \\
       6.0  &   -2.3406 &  -2.2177    &    -2.4872 &  -2.3579     &   -2.5854  & -2.4325 \\
       8.0  &   -2.4278 &  -2.6183    &    -2.5332 &  -2.6378     &   -2.5798  & -2.5656 \\
      10.0  &   -2.4208 &  -2.6528    &    -2.5176 &  -2.6391     &   -2.5444  & -2.5308 \\
      12.0  &   -2.4100 &  -2.6948    &    -2.4966 &  -2.6442     &   -2.5071  & -2.5007 \\ 
      14.0  &   -2.3771 &  -2.6785    &    -2.4571 &  -2.6080     &   -2.4589  & -2.4475 \\
      16.0  &   -2.3326 &  -2.6177    &    -2.4095 &  -2.5413     &   -2.4071  & -2.3775 \\
      18.0  &   -2.3245 &  -2.5542    &    -2.3918 &  -2.4744     &   -2.3774  & -2.3101 \\
      20.0  &   -2.3590 &  -2.4959    &    -2.4081 &  -2.4135     &   -2.3730  & -2.2492 \\ \hline
5.0 (cont)  &   -2.6522 &  -2.5524    &    -2.7447 &  -2.6072     &   -2.7865  & -2.6166 \\ \bottomrule
\enddata
\end{deluxetable}

\begin{deluxetable}{l c c | c c | c c}
\tabletypesize{\scriptsize}
\tablewidth{0pc}
\tablecaption{$\beta_{18}$ in Starburst99 Models with ROT Evolutionary Tracks, $f_{\rm esc}=0$} 
\tablecolumns{7}
\label{tab:v40}
\tablehead{
\colhead{Age}
& \multicolumn{2}{c}{IMF = 1.3,1.3}
& \multicolumn{2}{c}{IMF = 1.3,2.3}
& \multicolumn{2}{c}{IMF = 1.3,3.3} \\ \cline{2-3} \cline{4-5} \cline{6-7}
\colhead{(Myr)}
&\colhead{0.002}
&\colhead{0.014}
&\colhead{0.002}
&\colhead{0.014}
&\colhead{0.002}
&\colhead{0.014}
}
\startdata
       0.0  &   -2.4215  & -2.5341    &    -2.5448  & -2.6020    &    -2.6486  & -2.5968 \\
       2.0  &   -2.4508  & -2.5753    &    -2.5539  & -2.6382    &    -2.6527  & -2.6316 \\
       4.0  &   -2.2227  & -2.2496    &    -2.3848  & -2.3583    &    -2.5619  & -2.4609 \\
       6.0  &   -2.1270  & -2.0764    &    -2.3188  & -2.2163    &    -2.4825  & -2.3388 \\
       8.0  &   -2.1814  & -2.3088    &    -2.3377  & -2.3892    &    -2.4586  & -2.4229 \\
      10.0  &   -2.3127  & -2.6991    &    -2.4309  & -2.6865    &    -2.4946  & -2.5940 \\
      12.0  &   -2.4371  & -2.5492    &    -2.5180  & -2.5628    &    -2.5330  & -2.4866 \\
      14.0  &   -2.4955  & -2.6723    &    -2.5533  & -2.6280    &    -2.5402  & -2.4979 \\
      16.0  &   -2.4309  & -2.6436    &    -2.4918  & -2.5861    &    -2.4814  & -2.4441 \\
      18.0  &   -2.3848  & -2.5977    &    -2.4446  & -2.5323    &    -2.4325  & -2.3838 \\
      20.0  &   -2.3579  & -2.5487    &    -2.4131  & -2.4780    &    -2.3949  & -2.3260 \\ \hline
5.0 (cont)  &   -2.3655  & -2.3248    &    -2.4828  & -2.4522    &    -2.6107  & -2.5318 \\ \bottomrule
\enddata
\end{deluxetable}

\begin{deluxetable}{l c c | c c | c c}
\tabletypesize{\scriptsize}
\tablewidth{0pc}
\tablecaption{$\beta_{18}$ in Starburst99 Models with ROT Evolutionary Tracks, $f_{\rm esc}=100$} \tablecolumns{7}
\label{tab:v40}
\tablehead{
\colhead{Age}
& \multicolumn{2}{c}{IMF = 1.3,1.3}
& \multicolumn{2}{c}{IMF = 1.3,2.3}
& \multicolumn{2}{c}{IMF = 1.3,3.3} \\ \cline{2-3} \cline{4-5} \cline{6-7}
\colhead{(Myr)}
&\colhead{0.002}
&\colhead{0.014}
&\colhead{0.002}
&\colhead{0.014}
&\colhead{0.002}
&\colhead{0.014}
}
\startdata
       0.0  &   -3.0191 &  -3.0324     &   -3.0043 &  -2.9739    &    -2.8775 &  -2.7874 \\
       2.0  &   -3.0005 &  -3.0328     &   -2.9970 &  -2.9822    &    -2.8940 &  -2.8166 \\
       4.0  &   -2.3642 &  -2.5496     &   -2.5529 &  -2.5959    &    -2.7096 &  -2.6085 \\
       6.0  &   -2.2135 &  -2.1978     &   -2.4089 &  -2.3166    &    -2.5574 &  -2.4081 \\
       8.0  &   -2.2183 &  -2.3523     &   -2.3755 &  -2.4243    &    -2.4906 &  -2.4473 \\
      10.0  &   -2.3295 &  -2.7125     &   -2.4474 &  -2.6972    &    -2.5084 &  -2.6015 \\
      12.0  &   -2.4456 &  -2.5527     &   -2.5263 &  -2.5660    &    -2.5398 &  -2.4891 \\
      14.0  &   -2.5003 &  -2.6745     &   -2.5579 &  -2.6300    &    -2.5439 &  -2.4994 \\
      16.0  &   -2.4337 &  -2.6449     &   -2.4945 &  -2.5873    &    -2.4837 &  -2.4449 \\
      18.0  &   -2.3866 &  -2.5986     &   -2.4464 &  -2.5331    &    -2.4340 &  -2.3843 \\
      20.0  &   -2.3593 &  -2.5493     &   -2.4145 &  -2.4786    &    -2.3960 &  -2.3265 \\ \hline
5.0 (cont)  &   -2.7260 &  -2.6853     &   -2.7964 &  -2.7344    &    -2.8079 &  -2.6949 \\ \bottomrule
\enddata
\end{deluxetable}

\end{document}